\documentclass[aps,prd,onecolumn,superscriptaddress,showpacs,floatfix]{revtex4-2}%

\usepackage{graphicx}
\usepackage{tabularx}
\usepackage{bm}
\usepackage{latexsym}
\usepackage{epsf}
\usepackage{rotating}
\usepackage{epsfig,graphics,rotate,color}
\usepackage{wrapfig}
\usepackage{amssymb}
\usepackage{amsmath}
\usepackage{amsfonts}
\usepackage{subfigure}
\usepackage{array,hhline,dcolumn}%
\usepackage[normalem]{ulem}
\usepackage{color}
\usepackage{wasysym}
\usepackage[colorlinks=true,allcolors=blue]{hyperref}

\begin{document}
\bibliographystyle{apsrev4-1}

\title{The Width of a Beta-decay-induced Antineutrino Wavepacket}

\author{B.J.P. Jones}
\affiliation{
University of Texas at Arlington, Arlington, Texas 76019, USA
}

\author{E. Marzec}
\affiliation{
University of Michigan, Ann Arbor, Michigan 48109, USA
}

\author{J. Spitz}
\affiliation{
University of Michigan, Ann Arbor, Michigan 48109, USA
}

\begin{abstract}
The time evolution of a neutrino is dependent on its initial properties at creation including flavor, energy, and wavepacket size. There exists no solid theoretical prediction for the latter property in the context of nuclear beta decay, despite the importance of this process for the past, present, and future of neutrino experimentation. In this paper, we provide a quantitative prediction for the size of a beta-decay-induced electron antineutrino wavepacket by treating the  parent nucleus decaying to an entangled antineutrino-recoil system using the formalism of open quantum systems. Of central importance is the delocalization scale of the parent particle. We construct a systematic description of the hierarchy of localizing entanglements that provides an unambiguous statement of the relevant localization scale, found to be closely related to the diameter of the parent nucleus (e.g. $\sim$5-6~fm for beta-decaying fission daughters) and as low as the typical nucleon-nucleon correlation distance ($\sim$1~fm). Inside a nuclear reactor, for example, this translates to initial electron antineutrino wavepacket widths in the $\sigma_{\nu,x}\sim10\mathrm{-}400$~pm range for $E_{\overline{\nu}_e}>1.8$~MeV, with dependencies on decaying nucleus size, the emitted antineutrino energy, and the kinematics of the recoiling system. Wavepacket sizes in this envelope do not produce an observable effect on oscillation probability in foreseeable reactor experiments in the standard three-neutrino model, including JUNO which is expected to be sensitive to $\sigma_{\nu,x}\lesssim3$~pm. 


\end{abstract}
\maketitle

\section{Introduction}
Neutrinos are created as wavepackets with a width dependent on the production process, kinematics of the decay, and environment of the parent.  That the finite width of the wavepacket has implications for neutrino oscillation phenomenology has  been long recognized and widely studied theoretically~\cite{kayser2010testing,akhmedov2012neutrino,jones2015dynamical,coloma2018decoherence,coelho2017nonmaximal,gomes2019quantum,akhmedov2009paradoxes,giunti1998coherence,giunti2002neutrino,cohen2009disentangling,wu2010dynamics,wu2011neutrino}.  Interestingly, however, the width has so far eluded experimental measurement, and there are a number of neutrino production cases that lack a solid theoretical prediction for this characteristic size. In particular, no prediction exists for the earliest-known (anti)neutrino production process, nuclear beta decay. The naively relevant distance scales that may contribute to the width cover a wide range. In the case of antineutrinos coming from beta-decaying actinide fission daughters inside of a nuclear reactor, for example, relevant distances might include the inverse of the antineutrino energy~$\sim$0.1-0.7~pm, the inter-atomic spacing inside a Uranium-based nuclear reactor fuel rod~$\sim$0.1-1~nm, the diameter of the parent nucleus~$\sim$5-6~fm, the distance between correlated nucleon pairs~$\sim$1~fm, the size of a fuel rod~$\sim$10-100~cm, and the size of a reactor core~$\sim$1-10~m.

Despite the lack of measurement so far, experimental sensitivity to a realistic neutrino creation wavepacket width may be achieved in the future through its effect on neutrino oscillations. Considering a propagating neutrino as an evolving wavepacket, rather than as a single-momentum plane wave, results in the mass eigenstates separating in space due to their different group velocities. This process acts to suppress the nominally expected oscillation behavior in (e.g.) the three-neutrino mixing framework, and is dependent on distance traveled, energy, the mass splittings involved, and the initial wavepacket size.


The upcoming reactor-based JUNO experiment~\cite{JUNO:2015zny,JUNO_2022_physics_paper} has been shown to be highly sensitive to the wavepacket effect on oscillation observables, at the level of over an order of magnitude than has previously been achieved~\cite{deGouvea:2020hfl,JUNO:2021ydg,Marzec:2022mcz}. Considering a simplified model where each antineutrino has the same wavepacket width $\sigma_{\nu,x}$, JUNO is expected to be able to produce a two-sided constraint for $\sigma_{\nu,x}$ values as high as $3\times 10^{-12}$~m after 6~years of running~\cite{Marzec:2022mcz}, which can be compared to existing results from a phenomenological combination of Daya Bay, KamLAND, and RENO data of $\sigma_{\nu,x}>2.1\cdot 10^{-13}$~m at 90\%~CL~\cite{deGouvea:2021uvg} and $\sigma_{\nu,x}>1\cdot 10^{-13}$\,m at 95\% CL from a dedicated measurement with Daya Bay individually~\cite{DayaBay:2016ouy}. Accelerator-based and atmospheric-based experiments are wholly insensitive to the effect, and there is little hope for future realistic measurements~\cite{jones2015dynamical}. Notably, JUNO likely represents the only reasonable hope of measuring the wavepacket effect in the foreseeable future, across all mixing experiments, unless a higher mass splitting (e.g. at $\Delta m^2\sim$1~eV$^2$, as possibly indicated by Refs.~\cite{miniboone_new,lsnd3,source,Barinov:2021asz}) contributes to oscillations~\cite{Arguelles:2022bvt}, in which case the effect may be seen in experiments sensitive to this high-frequency mixing. The prospect of a wavepacket measurement is particularly exciting because it would not just represent an important new measurement in neutrino physics, it would also offer a fundamental test of quantum mechanics, validating predictions of the theory of measurement in open quantum systems where the generation of environmental entanglement leads to the emergence of classicality in quantum mechanics~\cite{zurek2003decoherence}. Therefore, before JUNO comes online in 2023, we consider it imperative to provide a grounded theoretical prediction for the characteristic range of widths of a beta-decay-induced antineutrino wavepacket, which immediately sets the expected oscillation spectrum observable.

In this paper, we produce an unambiguous and quantitative prediction for the range of wavepacket widths of an antineutrino resulting from nuclear beta decay using the density matrix formalism. This formalism provides a mathematical framework for handling the hierarchy of complicated entanglements that contribute to the width of the antineutrino at creation, including decoherence effects arising from the coupling of the decaying neutron to other nucleons in the nucleus, the confinement and momentum distribution of the decaying neutron, the nuclear interaction with its atomic electrons, interactions of the parent atom with other nearby atoms, etc. Robust treatment of all the relevant particles as members of an open quantum system allows us to make a prediction for the wavepacket width without ultimately relying on some assumed scale of localization for the system, as is required in many other treatments to date.  We predict that the range of the typical parent localization scales that sets the antineutrino wavepacket width is related to the nucleon-nucleon correlation distance and the nuclear diameter, and that the antineutrinos from a nuclear reactor emerge with a range of wavepacket sizes, with each one depending on the energy of the antineutrino, identity of the decaying nucleus, and kinematics of the entangled recoiling system.  An approximation is used for the nuclear wave function of the decaying parent that includes representative scales for the nuclear size and distance of nucleon-nucleon correlations, though our formalism in principle allows the use of an \textit{ab-initio} wave function, a potentially  viable prospect for ultra-light nuclei where these can be calculated using existing techniques. Unfortunately, based on this prediction, we expect JUNO to be insensitive to the effect of wavepacket width on oscillations.

We begin Sec.~\ref{sec:Coherence-properties-of} with a general introduction to decoherence in quantum systems and the implementation of density matrix theory towards making predictions about the probabilistic properties of such systems coupled to an environment.  In Sec.~\ref{sec:Outline-of-the}, we begin the neutrino-specific discussion by showing how a neutrino wavepacket affects oscillation probability and then derive the relationship between parent localization scale and neutrino coherence distance in Sec.~\ref{sec:Obtaining-the-neutrino}. With that relationship in hand, we predict the relevant range of localization scales of the parent particle in Sec.~\ref{sec:Finding-the-delocalization} and present the experimental implications of these results and outlook/conclusions in Sec.~\ref{Implications} and \ref{Conclusions}.
\section{Coherence properties of open quantum systems \label{sec:Coherence-properties-of}}

The basic principle of decoherence theory
is that the interaction between an open quantum system and its environment generates
entanglement, which suppresses coherence~\cite{zurek2003decoherence,joos2013decoherence,schlosshauer2005decoherence}. The encoding of quantum
information about the system into the environment in this way serves
as a ``measurement'', regardless of whether the information is ultimately
accessed by an experimenter. In the Young's two slit experiment,
for example, if any degree of freedom in the environment encodes sufficient
information to determine which slit was traversed then the interference
pattern will vanish - it is not necessary to access this information
experimentally. This principle has been convincingly demonstrated
in the laboratory using Talbot Lau interferometry~\cite{hackermuller2003decoherence}. The emergence
of stable effective wavepacket widths of systems in different environments
where the entanglement is generated by scattering processes has been
described in Ref.~\cite{tegmark1993apparent}. This principle was used to calculate the
width of neutrino wavepackets in pion decay-in-flight beams~\cite{jones2015dynamical}.
It will allow us to obtain the width of the neutrino wavepacket in
a nuclear decay system without ambiguity.

Because treatments using wave functions rather than density matrices
are more common in the neutrino physics literature, we will first
briefly review the use of reduced density matrices for treating
entangled bipartite systems, and their connection to the wave-function description. Readers familiar with density matrices
may wish to skip this section and proceed directly to Sec. \ref{sec:Outline-of-the}.

A density matrix provides the most
general description of a quantum system~\cite{nielsen2002quantum}. A system with a known quantum
state $|\Psi\rangle$ is called a pure state and has a density
matrix $\rho_{\mathrm{Pure}}$ defined by
\begin{equation}
\rho_{\mathrm{Pure}}=|\Psi\rangle\langle\Psi|,\label{eq:PureState}
\end{equation}
whereas a probabilistic distribution of pure states $|\Psi_{i}\rangle$
with probabilities $P_{i}$ is represented by a mixed state density
matrix $\rho_{\mathrm{Mixed}}$,
\begin{equation}
\rho_{\mathrm{Mixed}}=\sum_{i}P_{i}|\Psi_{i}\rangle\langle\Psi_{i}|.
\end{equation}
The decomposition of $\rho_{\mathrm{Mixed}}$ into $|\Psi_{i}\rangle$
with associated $P_{i}$ is always possible, but not unique. Measurements
are made on $\rho$ projectively. For example, if we measure some
observable $A$ the probability of finding outcome $a$ from among the eigenvalues of $A$ is given by
\begin{equation}
P_{a}=\langle a|\rho|a\rangle.\label{eq:POVM}
\end{equation}
The probabilities for any measurement that can be made on a quantum
system can be calculated using Eq. \ref{eq:POVM}. A particular class
of measurements that are of interest to us are those that probe a subsystem
$s$ entangled with an environment $\epsilon$, but without reference
to any information that is held in the environment. In this case,
the state can be written as a sum of products on the two Hilbert
spaces, as
\begin{equation}
|\Psi\rangle=\sum_{ij}c_{ij}|s_{i}\rangle\otimes|\epsilon_{j}\rangle.\label{eq:NonDiag}
\end{equation}
When constructing measurements on $s$ we must account for the fact
that some information about $s$ is encoded in $\epsilon$ via entanglement.
Consider that an external party makes a measurement $M$ on the environment
$\epsilon$ that finds result $\lambda_{k}$, for $M|\lambda_{k}\rangle=\lambda_{k}|\lambda_{k}\rangle$.
This would collapse the state of $s$ to $|\psi_{k}\rangle$
with probability $P_{k}$, leaving the density matrix of system
$s$ alone in a mixed state:
\begin{equation}
\rho_{s}=\sum_{k}P_{k}|\psi_{k}\rangle\langle\psi_{k}|=\sum_{k}\langle\lambda_{k}|\rho|\lambda_{k}\rangle\equiv\mathrm{Tr}_{\epsilon}[\rho],
\end{equation}
where the final equality serves to define the partial trace over $\epsilon$.
This object is called the reduced density matrix $\rho_{s}$. While
it is clear that $\rho_{s}$ represents the density matrix
of system $s$ after the measurement $M$ has been made on $\epsilon$,
its true power becomes apparent when we imagine some other measurement
$M'$ were made on $\epsilon$, since as students of only system
$s$ we do not know which measurement was made or what its outcomes
were. There are now a different set of possible outcomes $|\lambda'_{k}\rangle$
related to the original $|\lambda_{k}\rangle$ by a unitary transformation,
$|\lambda_{k}'\rangle=U_{ki}|\lambda_{i}\rangle$. The new reduced
density matrix is $\rho'_{s}$:
\begin{equation}
\rho'_{s}=\sum_{k}\langle\lambda_{k}|U^{\dagger}|\psi\rangle\langle\psi|U|\lambda_{k}\rangle.
\end{equation}
Inserting two complete sets of states into this expression we find:
\begin{eqnarray}
\rho'_{s}&=&\sum_{kab}\langle\lambda_{k}|U^{\dagger}|\lambda_{a}\rangle\langle\lambda_{a}|\psi\rangle\langle\psi|\lambda_{b}\rangle\langle\lambda_{b}|U|\lambda_{k}\rangle \\
&=&\sum_{kab}U_{bk}U_{ka}^{\dagger}\langle\lambda_{a}|\psi\rangle\langle\psi|\lambda_{b}\rangle=\rho_{s}.
\end{eqnarray}
The reduced density matrix $\rho_{s}$ is thus independent of the
actual measurement performed on $\epsilon$. Since any measurement
made on $s$ alone can be treated using only $\rho_{s}$, we conclude that
it does not matter which observable in $\epsilon$ was measured, or
even whether one was measured at all, for predicting probabilities at $s$. The probabilities of
measurements made on $s$, when summed over possible outcomes from $\epsilon$, are independent of what was measured at $\epsilon$. This is the principle that protects causality in the
presence of spooky action-at-a-distance in the original EPR thought
experiment, while at the same time allowing for non-trivial quantum correlations between the two subsystems~\cite{bell1964einstein}. 

While the measurements made on $\epsilon$ do not influence the probabilities found at $s$ so long as the experimenter at $s$ ignores their outcomes, the entanglement between the two subsystems does have important implications since it serves to limit coherence within $s$. To see
how this occurs in practice, let us construct an illustrative toy
example. We will consider two possible quantum states, about which we will ask two specific questions. First, we will obtain the probability distribution
for measuring the position of a particle in $s$ and finding
it at 1D position $x$,
\begin{equation}
P(x)=\langle x|\rho_{s}|x\rangle.
\end{equation}
Second, we will consider the result of feeding this state into a two
slit interferometer with slits at $x_A$ and $x_B$ in the $x$ plane. The probability of finding a particle $L$ at the detection plane (which is not in the $x$ plane) can be written at $Q(L)$,
\begin{equation}
Q(L)=\langle{\cal I}|\rho_s|{\cal I\rangle}\quad\quad|{\cal I}\rangle=\left(|x_{A}\rangle e^{i\theta_{A}(L)}+|x_{B}\rangle e^{i\theta_{B}(L)}\right),
\end{equation}
with $\theta_{A}(L)$ and $\theta_{B}(L)$ representing the phases accumulated
at position $L$ by particles passing through slit $A$ or $B$.

\begin{figure}
\begin{centering}
\includegraphics[width=0.99\columnwidth]{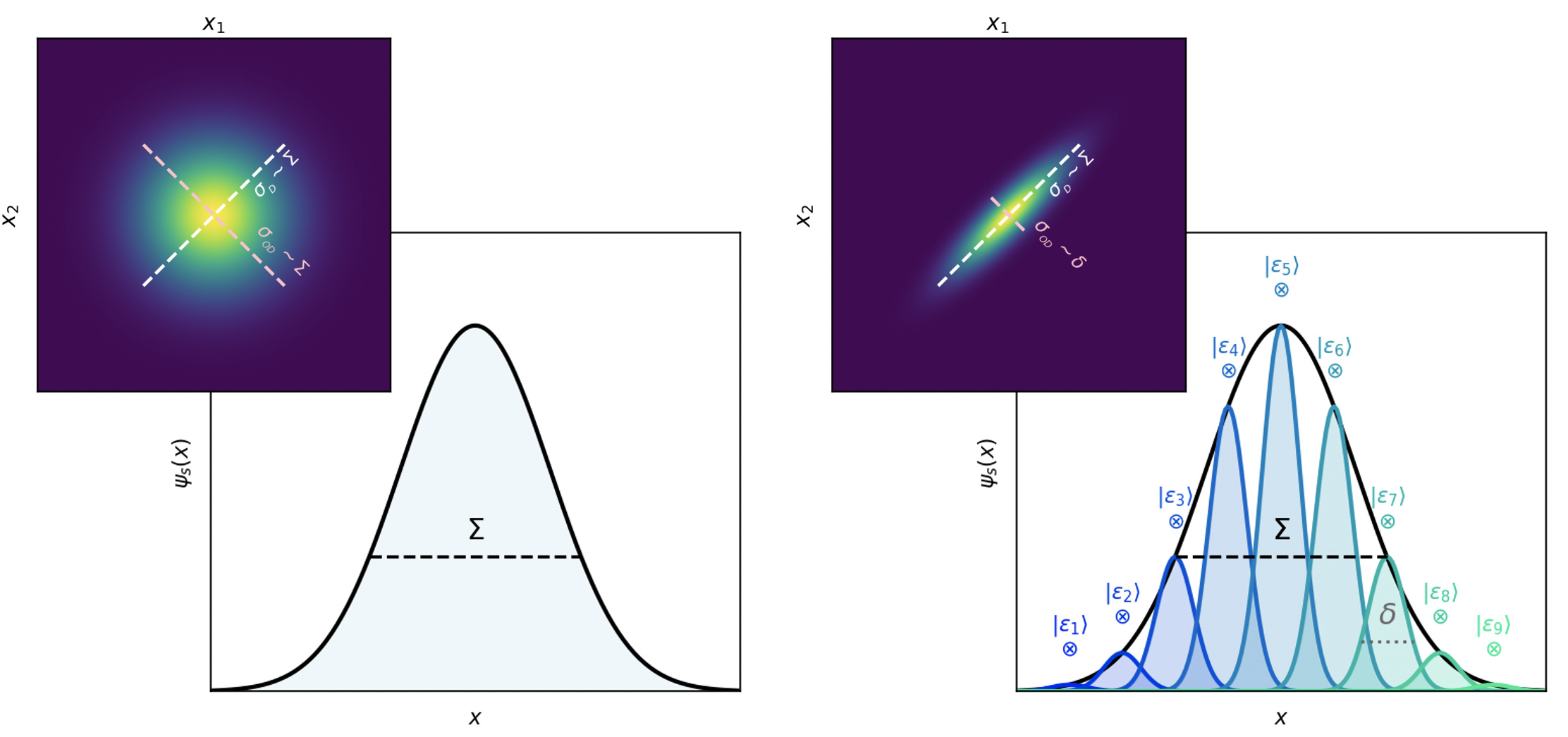}
\par\end{centering}
\caption{The reduced density matrix (colored panels) and wave functions (plots)
for the unentangled (left) and environmentally entangled (right) system
states. While the uncertainty on the position of the system is always
$\sigma_{D}\sim\Sigma$, the scale over which coherent quantum interference
effects can be observed is limited by the precision with which the environment
encodes the system position, $\sigma_{OD}\sim\delta$. \label{fig:The-reduced-density}}

\end{figure}

The first state we will consider is an isolated system that is unentangled
with any external degrees of freedom. In this case the state vector
and reduced density matrix are defined by a single width parameter
$\Sigma$,
\begin{eqnarray}
|\psi\rangle&=&\mathcal{N}\int dx\exp\left(-\frac{(x-x_{0})^{2}}{4\Sigma^{2}}\right)|x\rangle,\\
\rho_{s}&=&\mathcal{N}\int dx_{1}dx_{2}\exp\left(-\frac{(x_{1}-x_{0})^{2}+(x_{2}-x_{0})^{2}}{4\Sigma^{2}}\right)|x_{1}\rangle\langle x_{2}|.
\end{eqnarray}
The reduced density matrix and state vector are shown in Fig. \ref{fig:The-reduced-density},
left. Here and throughout this paper we will collect normalization
constants into a generic $\mathcal{N}$, which we can restore at the end of
the calculation should we so choose. We now ask our two prescribed
questions. The probability for finding the system at $x$ is what
we would expect it to be:
\begin{equation}
P(x)=\frac{1}{\left(2\pi\Sigma^{2}\right)^{1/2}}\exp\left(-\frac{(x-x_{0})^{2}}{2\Sigma^{2}}\right),
\end{equation}
and the result of the two slit experiment is
\begin{equation}
Q(L)=Q_{x_A}(L)+Q_{x_B}(L)+\frac{1}{\left(2\pi\Sigma^{2}\right)^{1/2}}\exp\left(-\frac{(x_{A}-x_{0})^{2}+(x_{B}-x_{0})^{2}}{4\Sigma^{2}}\right)\cos\left[\theta_{A}(L)-\theta_{B}(L)\right].
\end{equation}
Here $Q(L)$ is the probability of detecting a particle at coordinate $L$ on the detection plane, and $Q_{x_A}(L)$ and $Q_{x_B}(L)$ represent the probabilities that would be obtained if only one slit were open and the other were closed.  The interference experiment shows non-negligible interference effects
so long as both $x_{A}$ and $x_{B}$ are within the region of space
where the wave function is non-negligible, i.e. $\left(x_{A}-x_{0}\right)^{2}\apprle\Sigma^{2},\left(x_{B}-x_{0}\right)^{2}\apprle\Sigma^{2}$. 

Next we consider a scenario where we have a system entangled into
an environment. A degree of freedom in the environment $y_{\epsilon}$
is entangled with the system and encodes the system position to precision
$\delta$, but the range of allowed values of both $y_{\epsilon}$ and $x$ is
broad. A representative state vector and resultant reduced density matrix is
\begin{eqnarray}
|\Psi\rangle&=&\mathcal{N}\int dx\,dy_\epsilon\exp\left(-\frac{\left(y_{\epsilon}-x\right)^{2}}{4\delta^{2}}\right)\exp\left(-\frac{(x-x_0)^{2}}{4\Sigma^{2}}\right)|x\rangle\otimes|y_{\epsilon}\rangle,\quad\quad\delta\ll\Sigma.\\
\rho_s&=&\mathcal{N}\int dx_{1}dx_{2}\exp\left(-\frac{(x_{1}-x_{2})^{2}}{8\Delta_{OD}^{2}}\right)\exp\left(-\frac{(x_1-x_0)^2+(x_2-x_0)^2}{4\Delta_{D}^{2}}\right)|x_{1}\rangle\langle x_{2}|,
\end{eqnarray}
where we have identified two important length parameters $\Delta_{D}$
and $\Delta_{OD}$ characterizing the on- and off-diagonal distance
scales,
\begin{equation}
\Delta_{D}^{2}=\Sigma^{2},\quad\quad\Delta_{OD}=\delta^{2}.
\end{equation}
The reduced density matrix and a schematic decomposition into system wave functions accompanying distinct environmental basis states is shown in Fig. \ref{fig:The-reduced-density},
right. We again ask our two prescribed questions. The position distribution, governed by the on-diagonal scale $\Delta_{D}=\Sigma$, is found
to be exactly the same as for the previous case,
\begin{equation}
P(x)=\frac{1}{\left(2\pi\Delta_{D}^{2}\right)^{1/2}}\exp\left(-\frac{(x-x_{0})^{2}}{2\Delta_{D}^{2}}\right).
\end{equation}
On the other hand, the degree of interference in the interferometry
experiment is now limited by a further requirement, that $x_{A}$
and $x_{B}$ not be separated further than the entanglement scale
$\Delta_{OD}=\delta$.
\begin{equation}
Q(L)=Q_{x_A}(L)+Q_{x_B}(L)+\mathcal{N} \exp\left(-\frac{(x_{A}-x_{B})^{2}}{8\Delta_{OD}^{2}}\right)\exp\left(-\frac{(x_{A}-x_{0})^{2}+(x_{B}-x_{0})^{2}}{4\Delta_{D}^{2}}\right)\cos\left[\theta_{A}(L)-\theta_{B}(L)\right]
\end{equation}
If the slit separation is too large, information about which slit
the particle traversed is encoded through entanglement with the environment
- or speaking more loosely, the environment measures the system too
precisely to allow interference, and so entanglement suppresses coherence.
It is notable that this particular kind of ``measurement'' takes
place regardless of whether a human experimenter in fact chooses to
extract that information from the environment.

One way to interpret the two effective widths of the density matrix
is that $\Delta_{OD}$ represents the scale of quantum delocalization
whereas $\Delta_{D}$ represents an uncertainty due to classical ignorance.  Under this perspective, classicality has emerged from entanglement, a basic tenet of the theory of decoherence~\cite{joos2013decoherence}.
To bolster this interpretation we are free to decompose the reduced and mixed
density matrix $\rho_{s}$ as a probabilistic sum over state vectors,
each of which is quantum mechanically localized to precision $\Delta_{OD}$.
While we know the width of the wave function, we are classically ignorant
about where its center $X$ is with a precision of $\Delta_{D}$.
In the limit $\Delta_{OD}\ll\Delta_{D}$ we may decompose $\rho_{s}$ as
\begin{equation}
\rho_{s}=\int dX\,{\cal P}(X)\,|\psi_{X}\rangle\langle\psi_{X}|,\label{eq:RhoDecomp}
\end{equation}
\begin{equation}
|\psi_{X}\rangle=\mathcal{N}\int dx\exp\left[-\frac{(x-X)^{2}}{4\Delta_{OD}^{2}}\right]|x\rangle,\quad\quad{\cal P}(X)=\mathcal{N}\exp\left[-\frac{(X-x_0)^{2}}{2\Delta_{D}^{2}}\right].
\end{equation}
This further supports the interpretation that the limits of quantum
coherence are determined by the off-diagonal width $\Delta_{OD}$ which is itself encoded in $|\psi_X\rangle$; and at the end of the
calculation we can take a classical probabilistic sum over $X$ with
width $\Delta_{D}$. Because
the degree of classical ignorance is an emergent concept deriving
from entanglement with an unobserved subsystem, such calculations
are not burdened by any arbitrariness as to the location of the quantum/classical divide. Quantum mechanics is absolutely
prescriptive about the predictions of probabilities in such entangled
systems, as well as about the conditions in which the results of experiments
appear classical vs quantum mechanical. 

\section{Calculating the neutrino oscillation probability \label{sec:Outline-of-the}}

The principles of Sec. \ref{sec:Coherence-properties-of} will now
be applied to the neutrino oscillation system. We must
take care to distinguish between two kinds of uncertainty - the classical-like
uncertainty reflecting where in space we might expect to find a neutrino,
were we to make a position measurement; vs its off-diagonal uncertainty
which reflects the distance scale over which the wavepacket can exhibit
coherent interference, determined by the degree of entanglement with
external degrees of freedom. It is the latter that will limit coherence
by the effects of wavepacket separation. 

The reduced density matrix of the neutrino system contains both mass
state and configuration space information. We define 
\begin{equation}
\rho_{\nu}=\int dx_{1}dx_{2}\sum_{ij}\rho_{ij}(x_{1},x_{2},t)|x_{1};m_{i}\rangle\langle x_{2};m_{j}|.
\end{equation}
At different points in the calculation it will be convenient
to use either the momentum or position bases; these  choices will be distinguished by the arguments provided for the function $\rho_{ij}$.  

A neutrino oscillation experiment is a projective measurement of flavor
$\beta$ at some position $L$. The oscillation probability can
be evaluated as
\begin{equation}
P(\beta,L)=\langle\beta,L|\rho_{\nu}(t)|\beta,L\rangle,\quad\quad|\beta,L\rangle=\sum_{\beta i}U_{\beta,i}|m_{i},L\rangle.\label{eq:OscProbDef}
\end{equation}
This has a notably similar form to the two-slit interferometer example
considered in Sec. \ref{sec:Coherence-properties-of}, where the role
of the two slit positions is now played by the three distinct values
for $m_{i}$ within the three neutrino mixing paradigm. Because of environmental entanglement the neutrino reduced
density matrix will be a mixed state, and we are always
free to make the decomposition of Eq. \ref{eq:RhoDecomp} in terms
of pure states when it is helpful to do so. Writing the neutrino reduced density matrix as
\begin{equation}
\rho_{\nu}=\int dx_{0}\,{\cal{P}}(x_{0})\,|\psi_{\nu,x_{0}}\rangle\langle\psi_{\nu,x_{0}}|,\label{eq:RhoDecomp-1}
\end{equation}
and substituting this into Eq. \ref{eq:OscProbDef} shows that the probabilities
for each $x_{0}$ add incoherently,
\begin{equation}
P(\alpha,L)=\int dx_{0}\,{\cal P}(x_{0})\,P_{x_{0}}(\alpha,L),\quad\quad P_{x_{0}}(\beta,L)=\langle\beta,L|\psi_{\nu,x_{0}}\rangle\langle\psi_{\nu,x_{0}}|\beta,L\rangle.\label{eq:RhoDecomp-1-1}
\end{equation}
Thus all questions of coherence loss through wavepacket separation
are encoded in $|\psi_{\nu,x_{0}}\rangle$ which has width $\sigma_{\nu,x}=\Delta_{OD}$.
In order to understand neutrino coherence loss through wavepacket
separation, we must obtain $\rho_{\nu}$ and hence $\sigma_{\nu,x}$. 

The neutrinos in our system of interest are produced in the decay
of radioactive nuclei. These nuclei are already localized
within their environments, and the neutrino inherits these entanglements.
The reduced density matrix for the parent $\rho_{A}$ thus determines
the reduced density matrix for the neutrino, with entanglement against
the recoil system accounted for through the partial trace operation applied to the final state.
This calculation is performed in Sec. \ref{sec:Obtaining-the-neutrino}.
In terms of the neutrino width $\sigma_{\nu,x}$ we will find the
following neutrino oscillation probability,
\begin{equation}
P(\beta,L)=\mathcal{N}\sum_{ij}U_{\beta i}U_{\beta j}^{*}U_{\alpha i}U_{\alpha j}^{*}\exp\left[i\frac{\Delta m^{2}}{2E_{\nu}}L\right]\exp\left[-\frac{1}{32E_{\nu}^{4}}\left(\Delta m^{2}\right)^{2}\frac{L^{2}}{\sigma_{\nu,x}^{2}}\right]\exp\left[-\left(\frac{\Delta m^{2}}{2E}\right)^{2}\sigma_{\nu,x}^{2}\right],
\end{equation}
where $\sigma_{\nu,x}^{2}$ will be shown to be related to the parent
particle localization scale by
\begin{equation}
\sigma_{\nu,x}=\frac{2m_A^2}{M^2}\left(1+\frac{p_{A,0}}{\sqrt{m_A^{2}+p_{A,0}^{2}}}\right)^{-1}\sigma_{A,x},
\label{sigma_nu_equation}
\end{equation}
where $m_A$ is the parent mass and $M^2=m_A^2-m_X^2$, with $m_X$ as the invariant mass of the recoiling system. As is required by causality, any subsequent
interactions or detection of the entangled recoil after
emission will not influence the total oscillation probability. For
more discussion on this point, see  Refs.~\cite{jones2015dynamical,jones2022comment}.

The actual distance scale of the off-diagonal width of $\rho_{A}$
is obtained in Sec. \ref{sec:Finding-the-delocalization} by consideration of the hierarchy of entanglements experienced by the immediate precursor to the neutrino, and is determined to have a width to be of order the nuclear size. The implications of this are calculated for the reactor antineutrino case in Sec~\ref{Implications}. 

\section{From parent localization scale to neutrino coherence distance
\label{sec:Obtaining-the-neutrino}}

Let us first evaluate the neutrino wavepacket width $\sigma_{\nu,x}$
we would expect if a parent particle with wave function localized
to $\sigma_{A,x}$ decays to make a neutrino and some entangled recoiling
system that has definite invariant mass $m_{X}$. We can make the
decomposition of Eq.~\ref{eq:RhoDecomp} and consider the parent density
matrix as a probabilistic sum over such state vectors, where $\sigma_{A,x}=\Delta_{A,OD}$.
The initial state state vector of the parent then is
\begin{equation}
|\psi_{A,x_{0}}\rangle=\frac{1}{\left(2\pi\sigma_{A,x}^{2}\right)^{1/4}}\int dx\,\exp\left(-\frac{(x-x_{0})^{2}}{4\sigma_{A,x}^{2}}\right)e^{ip_{A,0}x}|x\rangle.
\end{equation}
We Fourier transform this to find the momentum representation, and
allow the parent to decay to a recoil system and a neutrino of flavor
$\alpha$, conserving momentum in the plane wave basis,
\begin{equation}
|\psi_{A}\rangle\rightarrow|\Psi\rangle=\sum_{i}\frac{U_{\alpha i}}{\left(2\pi\sigma_{A,p}^{2}\right)^{1/4}}\int dp_{A}\,\exp\left(-\frac{(p_{A}-p_{A,0})^{2}}{4\sigma_{A,p}^{2}}\right)e^{ip_A x_{0}}|p_{\nu}(p_A),m_{i}\rangle\otimes|p_{X}(p_A)\rangle.
\end{equation}
Here $p_{\nu}(p_A)$ and $p_{X}(p_A)$ are the neutrino and recoil system
emitted against a parent with initial momentum $p_{A}$ and mass $m_{A}$,
and $\sigma_{A,x}=1/2\sigma_{A,p}$, with natural units assumed throughout. We would like to express this
as an integral over neutrino momenta, which are related to the parent
momentum by
\begin{equation}
p_{A}=\frac{4m_{A}^{2}p_{\nu}^{2}-M^{4}}{4M^{2}p_{\nu}}.
\end{equation}
We can change the variables in the integral
by employing a suitable Jacobean for the transformation $J[p_{A}]$,
and since it is a function that varies slowly where the integral has support
we can evaluate it at $p_{A,0}$ and remove it from the integrand.  We also take x$_0$=0, keeping in mind we can always restore it later by spatially translating the final state by an amount $x_0$ after intermediate manipulations.
This leads us to the final state vector  expressed in terms
of an integral over neutrino momenta,
\begin{equation}
|\Psi\rangle=\frac{J^{-1}[p_{A,0}]}{\left(2\pi\sigma_{A,p}^{2}\right)^{1/4}}\sum_{i}\int dp_{\nu}\,\exp\left(-\frac{\left(p_{\nu}-\frac{M^{4}}{4m_{A}^{2}p_{\nu}}-\frac{M^{2}}{m_{A}^{2}}p_{A,0}\right){}^{2}}{4\frac{M^{4}}{m_{A}^{4}}\sigma_{A,p}^{2}}\right)|p_{\nu},m_{i}\rangle\otimes|p_{X}(p_{\nu})\rangle\label{eq:FullEqn}
\end{equation}
This distribution is not Gaussian. But it does have a peak at $p_{0}^{\nu}$
and $\pm1\sigma$ points at positions approximately $p_{0}^{\nu}\pm\sigma_{\nu,p}$
given by
\begin{equation}
p_{0}^{\nu}=\frac{M^{2}}{2m_A^{2}}\left(p_{A,0}+\sqrt{m_A^{2}+p_{A,0}^{2}}\right),\quad\quad\sigma_{\nu,p}=\frac{M^2}{2m_A^2}\left(1+\frac{p_{A,0}}{\sqrt{m_A^{2}+p_{A,0}^{2}}}\right)\sigma_{A,p}.\label{eq:DecayingWPWidth}
\end{equation}
The position-space widths are related by the inverse condition Eq.~\ref{sigma_nu_equation}, and the expression is valid when $\sigma_{p}\ll m,M$. As a good approximation
to Eq. \ref{eq:FullEqn} we can consider a Gaussian wave function
with mean $p_{0}^{\nu}$ and width $\sigma_{\nu,p}=\left(p_{1\sigma,\nu}-p_{0}^{\nu}\right)$,
as
\begin{equation}
|\Psi\rangle=\frac{1}{\left(2\pi\sigma_{\nu,p}^{2}\right)^{1/4}}\sum_{i}\int dp_{\nu}\,\exp\left(-\frac{\left(p_{i}-p_{0}^{\nu}\right){}^{2}}{4\sigma_{\nu,p}^{2}}\right)|p_{\nu},m_{i}\rangle\otimes|p_{X}(p_{\nu})\rangle.\label{eq:GaussApprox}
\end{equation}

\begin{figure}[t]
\begin{centering}
\includegraphics[width=0.49\columnwidth]{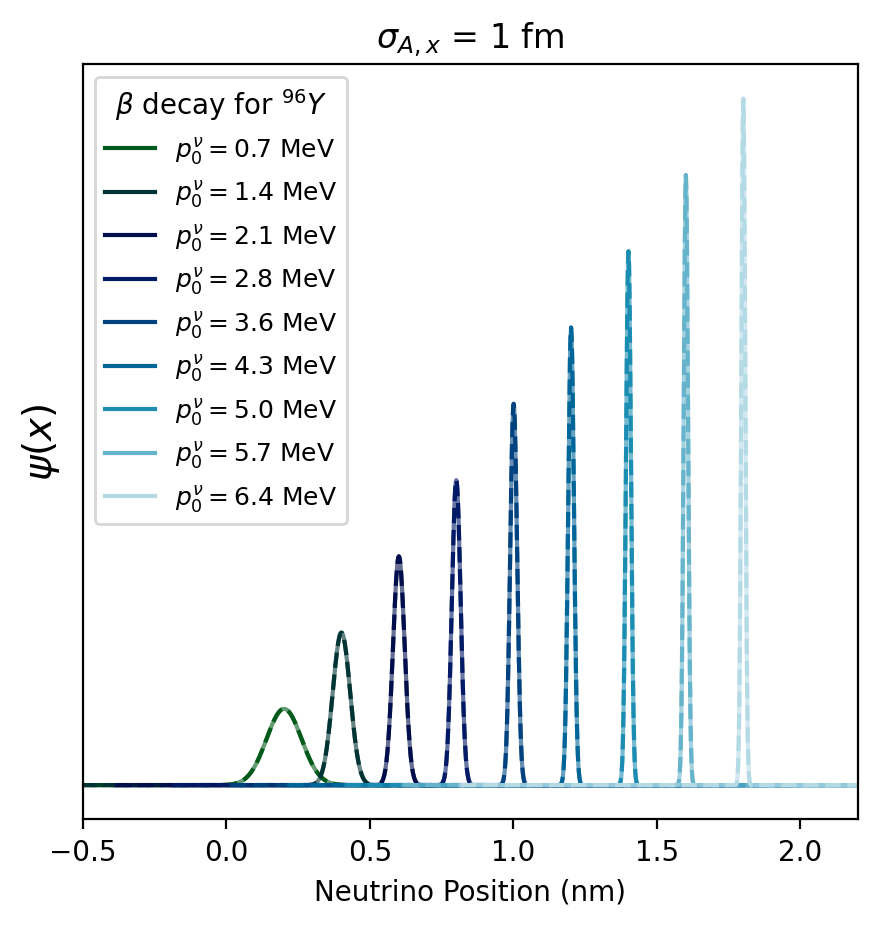}
\includegraphics[width=0.49\columnwidth]{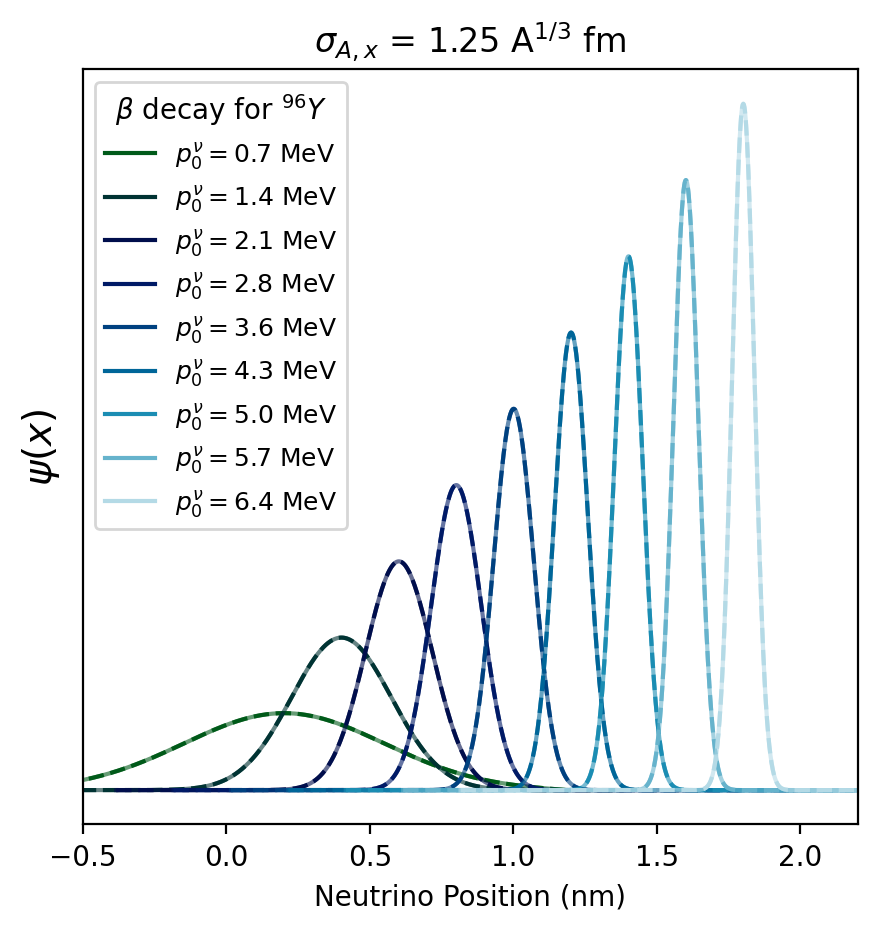}
\vspace{-.3cm}
\caption{Comparison of exact position-space wave function (dashed line) with Gaussian approximation (solid line) of the antineutrino emitted alongside a fixed recoil invariant mass of recoil system from the beta decay of $^{96}\mathrm{Y}$ to $^{96}\mathrm{Zr}_{\mathrm{g.s.}}$ ($Q=7.1$~MeV). The two panels show the cases where the width is limited by the typical nucleon short-range correlation scale (left) and nuclear size (right).}
\label{fig:wfcomparison}
\end{centering}
\end{figure}

Direct comparison of plots of Eq. \ref{eq:FullEqn} to Eq. \ref{eq:GaussApprox}
using relevant parameters to our problems of interest are shown in Fig.~\ref{fig:wfcomparison}. The dependence of $\sigma_{\nu,p}^{2}$ on $m_i$
is weak, and we can neglect it without significant consequences for
the calculation. We now form the total system density matrix, and
obtain the neutrino reduced density matrix by performing the partial
trace over the recoil,
\begin{equation}
\rho_{\nu}=\mathrm{Tr}{}_{X}[\rho]=\int dp_{X}\langle p_{X}|\rho|p_{X}\rangle.
\end{equation}
This is the correct treatment when the recoiling system has a definite
invariant mass, and hence there is a strict relationship between parent momentum and final state momenta for two body kinematics. If we consider a multi-particle recoiling
system, for example the recoiling nucleus and electron in beta decay, then there is no definite invariant mass.  Nevertheless, since final states with different invariant masses necessarily have orthogonal wave functions, we can form the neutrino reduced density matrix by a probabilistic sum over final state invariant masses. The effective two-body treatment developed thus far remains valid for each component in the sum over $m_X$,
\begin{equation}
\rho_{\nu}=\mathrm{Tr}{}_{X}[\rho]=\int dm_{X}{\cal P}(m_{X})\int dp_{X}\langle p_{X},m_{X}|\rho|p_{X},m_{X}\rangle.
\end{equation}
The probability distribution ${\cal P}(m_{X})$ is determined from well-understood beta
decay kinematics. For clarity we will omit this integral in the following discussion, but
it is understood that the derived expressions should be integrated over the $m_{X}$ distribution for the decay in the final step.

With the initial state neutrino reduced density matrix in hand, we
can apply time evolution in the typical way to find $\rho_{\nu}(t)$
\begin{equation}
\rho_{\nu}(t)=e^{-iHt}\rho_{\nu}(0)e^{iHt}.
\end{equation}
This gives us the time dependent form of the density matrix $\rho_{\nu}$
\begin{equation}
\rho_{\nu}=U_{\alpha i}U_{\alpha j}^{*}\int dp_{1}dp_{2}\rho_{\nu}^{ij}(p_{1},p_{2})|p_{1},m_{i}\rangle\langle p_{2},m_{j}|
\end{equation}
\begin{equation}
\rho_{\nu}^{ij}(p_{1},p_{2})\propto\exp\left(-\frac{\left(p_{1}-p_{0,i}^{\nu}\right){}^{2}}{4\sigma_{\nu,p}^{2}}-iE_{1}^{i}t\right)\exp\left(\frac{\left(p_{2}-p_{0,j}^{\nu}\right){}^{2}}{4\sigma_{\nu,p}^{2}}+iE_{2}^{j}t\right)\delta\left[p_{X}(p_{1},m_{i})-p_{X}(p_{2},m_{j})\right]\label{eq:FullDensMatrix}
\end{equation}
To obtain an oscillation probability we project onto a flavor state
$\beta$ at distance $L$:
\begin{eqnarray}
P(\beta,L)&=&U_{\beta i}U_{\beta j}^{*}\int dq_{1}dq_{2}e^{i(q_{2}-q_{1})L}\langle q_{1},m_{i}|\rho_{\nu}|q_{2},m_{j}\rangle \\
&=&\mathcal{N}\int dp_{1}dp_{2}\delta\left[p_{X}(p_{1},m_{i})-p_{X}(p_{2},m_{j})\right]e^{-\frac{\left(p_{1}-p_{0,i}^{\nu}\right){}^{2}}{4\sigma_{\nu,p}^{2}}-i\left(E_{1}^{i}t+p_{1}L\right)-\frac{\left(p_{2}-p_{0,j}^{\nu}\right){}^{2}}{4\sigma_{\nu,p}^{2}}+i\left(E_{2}^{j}t-p_{2}L\right)}.
\end{eqnarray}

We will now deal with the delta function on the left. First we make
the substitution,
\begin{equation}
\delta\left[p_{X}(p_{1},m_{i})-p_{X}(p_{2},m_{j})\right]\propto\delta(p_{1}-p_{2}+2\delta),
\end{equation}
where we have introduced the parameter 
\begin{equation}
\delta=\frac{1}{2}\left.\frac{dp_{\nu}}{dm_{\nu}^{2}}\right|_{p_{X}}\Delta m^{2}=\frac{1}{2}\left.\frac{dp_A}{dm_{\nu}^{2}}\right|_{p_{X}}\Delta m^{2},
\end{equation}
which is evaluated from two-body kinematics. 
We will next use the integral representation of the delta function,
with an important modification. The effect of this delta function
is to enforce ideal energy and momentum conservation in the decay.
But for a decaying particle of finite lifetime, energy and momentum
are only conserved approximately. Thus we use a slightly broadened
function, with $\Xi\propto1/\Gamma$ given by the particle decay width,
\begin{equation}
\delta(p_{1}-p_{2}+2\delta)=\int_{-\infty}^{\infty}dx\exp\left[i\left(p_{1}-p_{2}+2\delta\right)x\right]\rightarrow\int_{-\infty}^{\infty}dx\exp\left[i\left(p_{1}-p_{1}+2\delta\right)x+\frac{x^{2}}{\Xi^{2}}\right].
\end{equation}
For a long-lived particle as we consider here, we expect $\Xi\gg\sigma_{\nu,x}$.
Evaluating both integrals over $p_{1},p_{2}$ and then $x$, we ultimately
obtain
\begin{equation}
P(\beta,L)=\mathcal{N}\sum_{ij}U_{\beta i}U_{\beta j}^{*}U_{\alpha i}U_{\alpha j}^{*}\exp\left[i\left(L+\bar{v}t\right)\frac{\Delta m^{2}}{4E_{\nu}}\right]\exp\left[-\frac{(L-\bar{v}t)^{2}}{2\Xi^{2}}\right]\exp\left[-\frac{\Delta v^{2}t^{2}}{8\sigma_{\nu,x}^{2}}\right]\exp\left[-\left(\frac{\Delta m^{2}}{2E}\right)^{2}\sigma_{\nu,x}^{2}\right].
\end{equation}
We have introduced the mean velocity $\bar{v}=\left(v_{1}+v_{2}\right)/2$
and $\Delta v=v_{1}-v_{2}$ is the velocity difference. We have also
written this expression in terms of $\sigma_{\nu,x}=1/2\sigma_{\nu,p}$.
From left to right we see
\begin{enumerate}
\item The oscillation phase, which has the usual form;
\item A term that restricts detectability to around $L\sim\bar{v}t$ with
precision $\Xi$ dictated by the parent decay width. This accounts
for the mean expected arrival time of the wavepacket;
\item A term that suppresses coherence through wavepacket separation, dictated
by the neutrino localization scale $\sigma_{\nu,x}$. A careful evaluation
of this term is the main target of this work;
\item A term that smears oscillations if the parent particle is more delocalized
than an oscillation wavelength. This term will not be relevant in
most oscillation experiments, and any residual smearing from it would typically be accounted for in Monte Carlo simulations of neutrino emission across a finite source.
\end{enumerate}
To see the effect of wavepacket separation explicitly, it makes sense
to evaluate the whole expression near $L\sim\bar{v}t$ where the detection
probability for finding any neutrino is large. We can also use the
fact that $\Delta v=\frac{\Delta m^{2}}{2E_{\nu}^{2}}$ to find:
\begin{equation}
P(\beta,L)=\mathcal{N}\sum_{ij}U_{\beta i}U_{\beta j}^{*}U_{\alpha i}U_{\alpha j}^{*}\exp\left[i\frac{\Delta m^{2}}{2E_{\nu}}L\right]\exp\left[-\frac{1}{32E_{\nu}^{4}}\left(\Delta m^{2}\right)^{2}\frac{L^{2}}{\sigma_{\nu,x}^{2}}\right]\exp\left[-\left(\frac{\Delta m^{2}}{2E}\right)^{2}\sigma_{\nu,x}^{2}\right].
\label{osc_eq}
\end{equation}
It is notable that the wavepacket separation term obtained in this limit corresponds precisely to the one obtained through a treatment that ignores the entangled recoil altogether, for example Ref.~\cite{giunti1998coherence}. 

\section{The localization scale of the parent particle in radioactive decay \label{sec:Finding-the-delocalization}}

Generally speaking we cannot expect perfectly Gaussian density matrices.
We can, however, characterize the diagonal and off-diagonal widths
of whatever density matrix we have with the following two metrics,
\begin{equation}
\bar{\Delta}_{OD}^{2}(x)=\frac{\int dy\,y^{2}\rho_{1}(y,-y)}{\int dy\,\rho_{1}(y,y)},\quad\quad\bar{\Delta}_{D}^{2}(x)=\frac{\int dy\,y^{2}\rho_{1}(y,y)}{\int dy\,\rho_{1}(y,y)}.
\end{equation}
For a properly normalized density matrix, $\int dy\,\rho(y,y)=\mathrm{Tr}[\rho]=1$,
but we include a superfluous factor in the denominator to absorb the
arbitrary $\mathcal{N}$ factors that are attached to some of our expressions.
It is straightforward to verify that, given a Gaussian density matrix
the expected widths are recovered, $\bar{\Delta}_{OD}=\Delta_{OD}$
and $\bar{\Delta}_{D}=\Delta_{D}$.

Many treatments of neutrino oscillations require a degree of hand-waving
to determine the scale of $\bar{\Delta}_{OD}$. Here we carefully
consider the hierarchy of localizing effects in nuclear beta decay and show that this scale
can in fact be determined in a manner that is free from ambiguity, and it is found to be of nuclear dimensions. 

In a nuclear beta decay, electron antineutrinos are emitted by neutrons decaying inside
a nucleus through $n\rightarrow p+e^-+\overline{\nu}_e$. There are several effects that may
be intuitively argued to localize a decaying nucleon in space. A subset
of these include: 1) Localizing effect of the decaying nucleon inside
the nucleus, through interaction with other nucleons; 2) Localization
of a nucleus within an atom, through interaction with the atomic electrons;
3) Localization of an atom within a material, through electromagnetic
fields or scattering interactions; 4) Localizing effect of the recoiling
nucleus interacting with surrounding material, through scattering;
5) Quantum delocalization of the entire piece of material that the nucleus/atom is in. We may also
discuss, for example, 6) Delocalization of the planet Earth within
the solar system and 7) Delocalization of the solar system within the galaxy,
and an in principle infinite number of additional abstractions. In
order to convincingly assess which of these distance scales matter
for the problem, we must construct a schematic description of this
hierarchy of localizations and establish which
of them determine the off-diagonal width of the reduced density matrix
$\rho_{A}$.

First we may consider the nuclear state vector, which we assume has
been properly anti-symmetrized:
\begin{equation}
|\psi\rangle=\int dx^{N}\psi(x_{a},x_{b},x_{c}...)|x_{a}\rangle\otimes|x_{b}\rangle\otimes|x_{c}\rangle...\label{eq:NuclWaveFunc}
\end{equation}
This state accounts for the distribution of nucleons inside the nucleus,
including the details of their correlation and entanglement with one
another. While it is very difficult to calculate such wave functions
exactly, it is fair to say that they exist in principle and can be
expressed on the Hilbert space of $N$ nucleons, as in Eq. \ref{eq:NuclWaveFunc}.  Our convention here is chosen such that $\psi(x_{a},x_{b},x_{c}...)$ represents a state with center of mass at the origin.

Another degree of quantum uncertainty is the delocalization of the nucleus itself, which can be encoded in the center of mass
wave function $\phi(X)$. The overall state of the system is
then generated via convolution
\begin{equation}
|\Psi\rangle=\int dX\phi(X)e^{iX\hat{P}}|\psi\rangle,
\end{equation}
where $\hat{P}$ is the generator of translations $\hat{P}=i\hbar\sum_{\alpha}\frac{\partial}{\partial x_{\alpha}}$
such that
\begin{eqnarray}
|\Psi\rangle&=&\int dx^{N}\Psi(x_{a},x_{b}...)|x_{a}\rangle\otimes|x_{b}\rangle...\\
\Psi(x_{a},x_{b}...)&=&\int dX\phi(X)\psi(x_{a}-X,x_{b}-X...)\label{eq:PhiDef}.
\end{eqnarray}
To examine the key features of this problem without resorting to a complete
\textit{ab-initio} nuclear wave-function, we will consider the following illustrative
example wave function for the nucleus,
\begin{equation}
\psi(x_{a},x_{b},x_{c}...)=\left[\psi_{N}(x_{a})\psi_{N}(x_{b})\psi_{N}(x_{c})...\times\right]\cdot\left[\psi_{c}(x_{a}-x_{b})\psi_{c}(x_{c}-x_{b})...\times\right]\cdot\delta\left(\sum_{\alpha}x_\alpha\right).\label{eq:COMWF}
\end{equation}
This expression has three essential components; first, the single
particle wave-function $\psi_{N}$ which accounts for the distribution
of a given nucleon within the nucleus and is approximately of nuclear
size; second, a correlation function $\psi_{c}$ that establishes
how each nucleon is correlated to each other one; and third, a delta function that ensures that this state has center of mass at the origin. In principle this
wave function can be appropriately anti-symmetrized if we ensure that
the correlator is chosen such that $\psi_{c}(x_{a}-x_{b})=-\psi_{c}(x_{b}-x_{a})$,
though the demonstration that follows will not achieve
this in practice. A real nuclear wave function must also account for the
spin degrees of freedom that we have neglected. Finally, in a real nucleus there is a distribution of correlation scales with the lower limit fixed by the subset of nucleons in short-range pairs; the relevant distance can be assumed to be about 1~fm, with $\sim$20\% of nucleons in specific neutron-proton correlated states~\cite{2017RvMP...89d5002H}. While acknowledging these deficiencies, we can develop a useful degree of intuition for this system by considering a simplified toy nuclear
wave function dictated by two distance scales: the nuclear size $\sigma_{N}$, and the scale of nucleon-nucleon correlations $\sigma_{c}$, with
Gaussian distributions for both cases.  Naturally we expect $\sigma_{c}<\sigma_{N}$,
though at any time we may take $\sigma_{c}\rightarrow\infty$ to examine
the behavior of a system with uncorrelated nucleons. We therefore choose
\begin{equation}
\psi_{N}(x)=\frac{1}{(2\pi\sigma_{N}^{2})^{1/4}}\exp\left(-\frac{x^{2}}{4\sigma_{N}^{2}}\right),\quad\quad\psi_{c}(x-y)=\frac{1}{(2\pi\sigma_{c}^{2})^{1/4}}\exp\left(-\frac{\left(x-y\right)^{2}}{4\sigma_{c}^{2}}\right),
\end{equation}
such that
\begin{equation}
\psi(x_{a},x_{b},x_{c}...)=\mathcal{N}\exp\left[-\sum_{\alpha}\frac{x_{\alpha}^{2}}{4\sigma_{N}^{2}}-\sum_{\alpha,\beta<\alpha}\frac{\left(x_{\alpha}-x_{\beta}\right)^{2}}{4\sigma_{c}^{2}}\right]\delta\left(\sum_{\alpha}x_\alpha - NX \right).
\end{equation}
A third distance scale $\sigma_{M}$ is introduced
to represent delocalization of the nucleus itself in the material,
such that $\sigma_{M}>\sigma_{N}>\sigma_{c}$. Again choosing a Gaussian
function for illustrative purposes, we find
\begin{equation}
\phi(X)=\frac{1}{(2\pi\sigma_{M}^{2})^{1/4}}\exp\left(-\frac{X^{2}}{4\sigma_{M}^{2}}\right).
\end{equation}
The full wave function of the decaying system takes the form
\begin{equation}
\Psi(x_{a},x_{b}...)=\mathcal{N}\int dX\exp\left[-\frac{X^{2}}{4\sigma_{M}^{2}}-\sum_{\alpha}\frac{\left(x_{\alpha}-X\right)^{2}}{4\sigma_{N}^{2}}-\sum_{\alpha,\beta<\alpha}\frac{\left(x_{\alpha}-x_{\beta}\right)^{2}}{4\sigma_{c}^{2}}\right]\delta\left(\sum_{\alpha}x_\alpha - NX \right).
\end{equation}
The $X$ integral can
be performed using the delta function with the result that
\begin{equation}
\Psi(x_{a},x_{b}...)=\mathcal{N}\exp\left[-A\sum_{\alpha}x_{\alpha}^{2}+B\sum_{\alpha\neq\beta}x_{\alpha}x_{\beta}\right]
\end{equation}
\begin{equation}
A=\frac{1}{4N^{2}\sigma_{M}^{2}}+\left(\frac{N-1}{N}\right)\frac{1}{4\sigma_{N}^{2}}+\frac{N-1}{4\sigma_{c}^{2}},\quad\quad B=-\frac{1}{4N^{2}\sigma_{M}^{2}}+\frac{1}{4N\sigma_{N}^{2}}+\frac{1}{4\sigma_{c}^{2}}.
\end{equation}

This recipe can also be applied recursively to build onto the problem
any number of subsequent stages of delocalization - for example, of
nucleus within atom; atom within material; material within laboratory,
and so on.  Examining the system at the present level of recursion, however, will prove subsequent ones unnecessary. Constructing
the reduced density matrix for a single nucleon,
\begin{equation}
\rho_{1}(x'_{a},x_{a})=\int d^{N-1}x\,\Psi(x_{a},x_{b},x_{c}...)\Psi^{*}(x'_{a},x_{b},x_{c}...).\label{eq:DensMat-1}
\end{equation}
The effective off-diagonal width will be given by
\begin{equation}
\bar{\Delta}_{OD}^2(x)=\frac{1}{4A}=\frac{1}{4}\left[\frac{1}{4N^{2}\sigma_{M}^{2}}+\left(\frac{N-1}{N}\right)\frac{1}{4\sigma_{N}^{2}}+\frac{N-1}{4\sigma_{c}^{2}}\right]^{-1}.\label{eq:DeltaOD}
\end{equation}

This expression features contributions from all three distance scales
so far introduced into the problem and will tend to be dominated by whichever is smallest.  We can examine its behaviour in three illustrative regimes. First, we note that if $\sigma_M\rightarrow 0$, the localization distance will also tend towards zero,
\begin{equation}
\lim_{\sigma_{M}\rightarrow0}\bar{\Delta}_{OD}=\sigma_{M}N\rightarrow0.
\end{equation}
In this limit, the wave function is simply Eq.~\ref{eq:COMWF}.  The delta function forces the center of mass to zero, which means that if any one nucleon decays, the location of that decay can in principle be inferred by precisely measuring the positions of all the remaining nucleons.  This regime is not a physically relevant one in our systems of interest since it is very unlikely that the center of mass position of the nucleus will ever be localized more precisely than the nuclear size.  

Next we may consider a delocalized
nucleus, though with no appreciable nucleon-nucleon correlations.
This corresponds to the limit $\sigma_{M}\gg\sigma_{N}$, while $\sigma_{c}\rightarrow\infty$.
Here we find the relevant limit of Eq.~\ref{eq:DeltaOD} to be
\begin{equation}
\lim_{\sigma_{M}\gg\sigma_{N},\sigma\rightarrow\infty}\bar{\Delta}_{OD}=\sqrt{\frac{N}{N-1}}\sigma_{N}.
\end{equation}
This result too has an intuitive interpretation. Given a nucleus which
is totally localized ($\sigma_{M}=0$), the RMS distance $\lambda$ between
any given nucleon and the center of mass of the remaining $N-1$ nucleons
is given by
\begin{eqnarray}
\lambda^{2}&=&\left\langle \left(x_{1}-\frac{1}{N-1}\sum_{\alpha\neq1}x_{\alpha}\right)^{2}\right\rangle =\frac{1}{(2\pi\sigma_{N}^{2})^{N/2}}\int d^{N}x\,\left(\left(x_{1}-\frac{1}{N-1}\sum_{\alpha\neq1}x_{\alpha}\right)^{2}\right)\exp\left(-\frac{\sum_{\alpha}x_{\alpha}^{2}}{\sigma_{N}^{2}}\right)\\
&=&\langle x_{1}^{2}\rangle+\left(\frac{1}{N-1}\right)^{2}\sum_{\alpha}\langle x_{\alpha}^{2}\rangle\\
&=&\sigma_{N}^{2}\left(\frac{N}{N-1}\right).
\end{eqnarray}
When a nucleon in a totally delocalized nucleus decays, it leaves
behind the remaining $N-1$ nucleons; and these each encode the position
of the decayed nucleon through entanglement to a distance scale of
exactly what we have predicted. This statement runs counter to some
past discussions in the literature, which have imagined that if the
nucleus is very delocalized, for example if the nucleus is inside an atom which is vibrating within a material
lattice or travelling through a gas, then the coherent width of the
neutrino could be much larger than the nuclear size. It is not so,
because the remaining nucleons entangle its position with distance
scale determined by the nuclear diameter.

We finally include nucleon-nucleon correlations within the nuclear wave function
with some correlation scale $\sigma_{c}$ such that $\sigma_{c}<\sigma_{N}\ll\sigma_{M}$.
In the limit where the scale of these correlations is much
smaller than the size of the nucleus itself, we find
\begin{equation}
\lim_{\sigma_{M}\gg\sigma_{N}\gg\sigma_{c}}\bar{\Delta}_{OD}=\frac{\sigma_{c}}{\sqrt{N-1}}.
\label{sigmac_eq}
\end{equation}
The interpretation of this result is that when nucleons are correlated on scales much smaller than the size of the nucleus,
the nucleons may be considered to measure each other's positions to
a precision smaller than the nuclear diameter, determined instead by the correlation scale. Of course, were all beta emitters to decay to the ground state, no information could be encoded in the arrangement of nucleons within the daughter nucleus about the antineutrino emission position. However, among the beta decays creating antineutrinos in nuclear reactors (above the detection threshold of 1.8~MeV), around 80\% (70\%) lead to excited daughter nuclei in the final state, as provided by the OKLO software~\cite{oklo}, described in the next section. 
Thus, the resulting final state of these nucleons can indeed be considered to encode information about the decay position, and the sub-nuclear distance scale $\sigma_c$ corresponds to the approximate resolution of this information. The simplified wave function above presents only a single value for $\sigma_c$  resulting in a localization scale given by the cumulative result of $N-1$ independent measurements each with precision $\sigma_{c}$, scaling as $\sigma_{c}/\sqrt{N-1}$. In reality $\sigma_c$ is expected to follow a distribution of correlation scales, ranging from typical short-range separation of $\sim1$~fm between correlated pairs of neutrons and protons for about 20\% of nucleons~\cite{2017RvMP...89d5002H} and extending to larger values within the nuclear diameter, with the limit $\sigma_c\rightarrow\infty$ corresponding to totally uncorrelated nucleons.  We leave a rigorous study of this distribution for future work, but conclude that the typical distance between correlated pairs of nucleons sets the lower limit of the relevant distance scales in the problem, and the nuclear diameter sets the upper limit.   Since no protons or neutrons are left with internal excitation following beta decay, degrees of freedom smaller than the nucleon correlation scale cannot be considered to encode localizing information about the antineutrino.

To summarize this section, we have shown that the primary scale
of localization that determines the emission coherence of neutrinos
from nuclear decay is at most the
nuclear size, with only an order-1 numerical prefactor $\sqrt{N/N-1}$. This is because the position of each nucleon
is always entangled with the position of the remainder of the nucleus,
even if the nucleus is itself delocalized.  The primary scale
of localization can be as small as that of the characteristic distance between nucleon-nucleon correlations within
the decaying nucleus. Since in practice there is a distribution of correlation lengths within the nucleus, we expect the true scenario to be a range of localization scales spanning from the shortest correlation distance to the nuclear size.  After accounting for the two effects above, the delocalization of the nucleus itself, including higher scales, is essentially irrelevant in determining the off-diagonal reduced density matrix width and hence the width of the antineutrino wavepacket.

To obtain a truly precise form for $\rho_{A}$, a real nuclear wave function should be used in Eq. \ref{eq:DensMat-1}. This can presently
only be calculated for very light nuclei~\cite{king2022ab,pastore2018neutrinoless} where convergent calculations exist, and this is of only marginal relevance for the problem of antineutrinos from nuclear reactors.  Nevertheless,
the derivations in this section demonstrate that the relevant scale for localization in the radioactive decay scenario lies in a range between the shortest nucleon-nucleon correlated distance to the nuclear diameter, already a specific enough range to make interesting and testable predictions for future reactor experiments.
We use this range of scales as input
to calculations of reactor antineutrino decoherence in the following Sec.~\ref{Implications}.

\section{Coherence loss for reactor neutrinos~\label{Implications}}
As presented in the previous sections, we predict that (1) the primary scale of localization that sets the coherence of neutrinos from nuclear decays ranges from the typical nucleon-nucleon correlation distance ($\sim1$~fm) to the diameter of the parent nucleus ($\sigma_{A,x}\approx1.25\cdot\mathrm{A}^{1/3}~\mathrm{fm}$, or about 5-6~fm for the A$\sim$80-140 beta decaying nuclei inside of a nuclear reactor); and (2) that the relationship between the parent wavepacket width and antineutrino wavepacket width is given by Eq.~\ref{sigma_nu_equation}. In this section, we examine the implications of these two predictions, in forming the antineutrino wavepacket width, on the experimental observable oscillation probability.

As shown in Eq.~\ref{sigma_nu_equation}, the mass of the parent fission product and invariant mass of the recoil system, dependent on the kinematics of each particular beta decay, set the relationship between the parent wavepacket width and the antineutrino wavepacket width. Towards finding this relationship on a decay-by-decay basis, we use the OKLO software toolkit~\cite{oklo} to simulate the hundreds of fission product beta decay branch contributions to a typical reactor-based electron antineutrino spectrum. OKLO uses inputs from various nuclear databases, including ENDF-B-VII.1~\cite{CHADWICK20112887}, JEFF-3.1.1, and JENDL-4.0~\cite{jendl} for providing the cumulative fission yields and ENSDF-6 data files for the beta feedings and properties of parent and daughter nuclei~\cite{Littlejohn:2018hqm}. While the software and database inputs have not been updated since 2015, we don't expect this fact to affect any results or conclusions presented here. We use a commercial Pressurized Water Reactor (PWR; $\mathrm{^{235}U}:\mathrm{^{238}U}:\mathrm{^{239}Pu}:\mathrm{^{241}Pu}=0.584:0.076:0.290:0.050$, consistent with the Daya Bay reactor complex), as a characteristic example of what can be expected from a low-enriched-uranium-based reactor experiment, noting that the fuel composition, including time evolution, will only weakly affect the results. 

\begin{figure}[t]
\begin{centering}
\includegraphics[width=10cm]{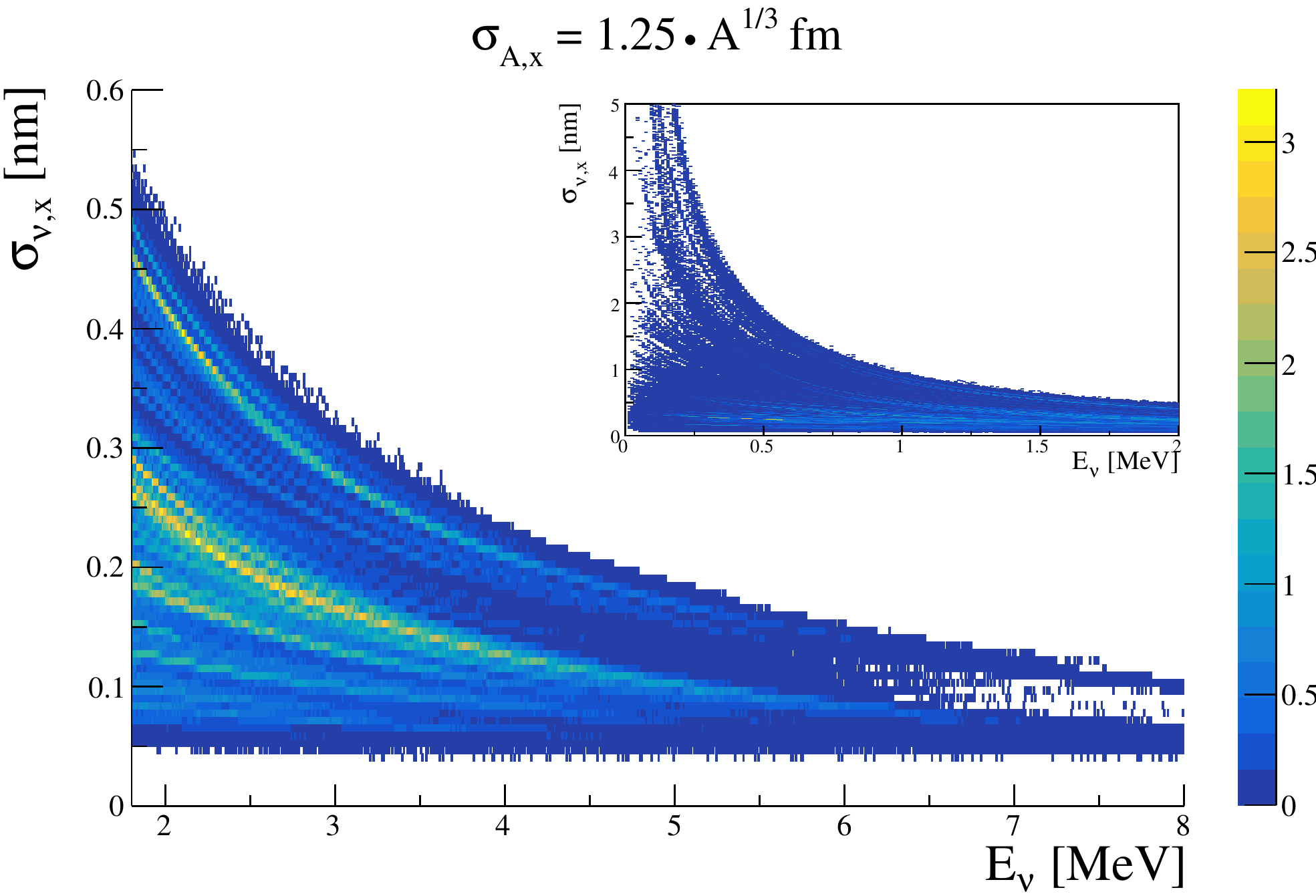}
\includegraphics[width=10cm]{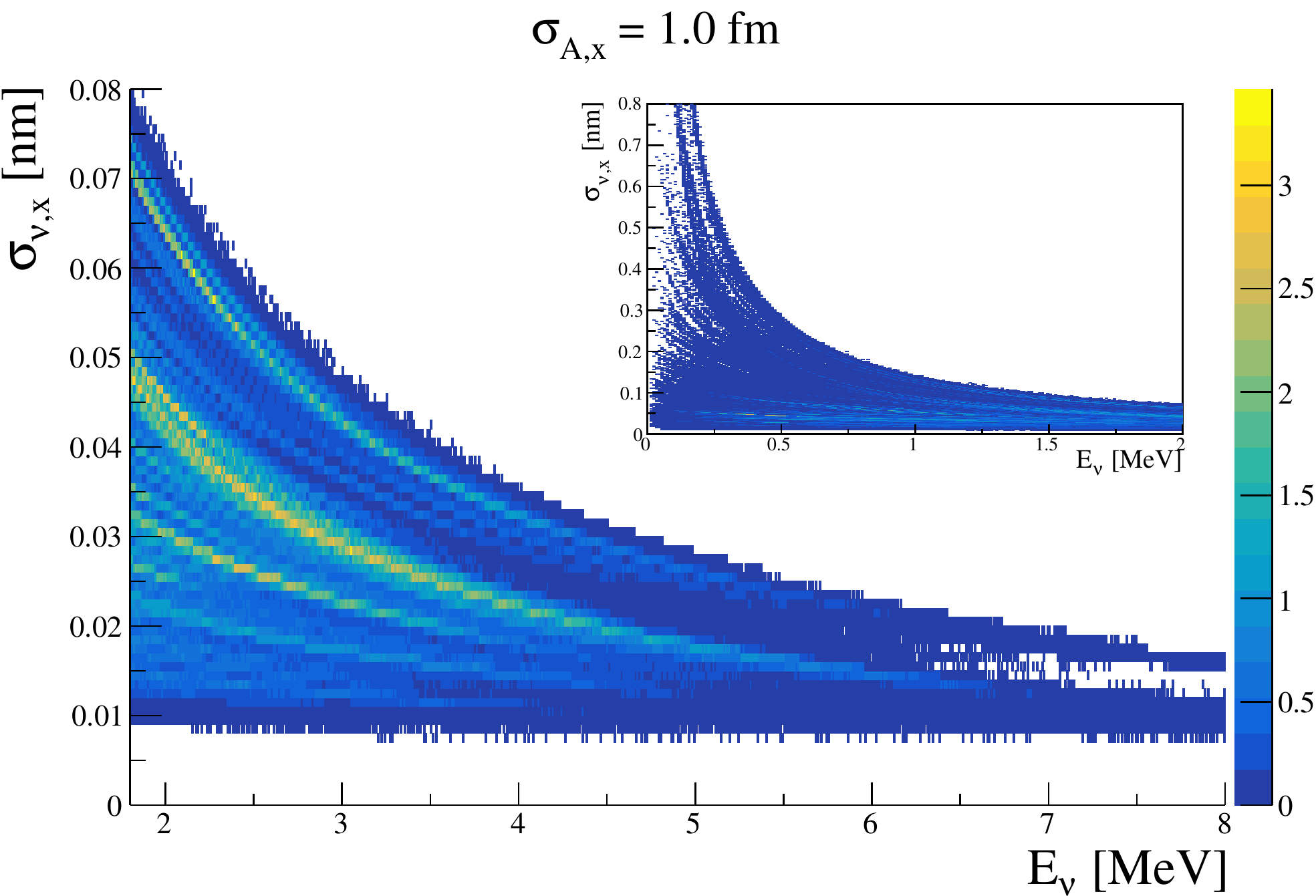}
\vspace{-.3cm}
\caption{Antineutrino wavepacket width vs energy for a beta-decay-parent width of $\sigma_{A,x}=1.25\cdot\mathrm{A}^{1/3}~\mathrm{fm}$ (top) and $\sigma_{A,x}=1$~fm (bottom). The visible bands correspond to individual beta decay branch contributions to the electron antineutrino spectrum. The main figures show energies relevant for $\overline{\nu}_e$ detection via IBD, with a threshold of 1.8~MeV, while the insets show the lower energy behavior.}
\label{sigmax_vs_energy_nuclear_width}
\end{centering}
\end{figure}

\begin{figure*}[ht]
\begin{center}
\includegraphics[width=0.45\textwidth]{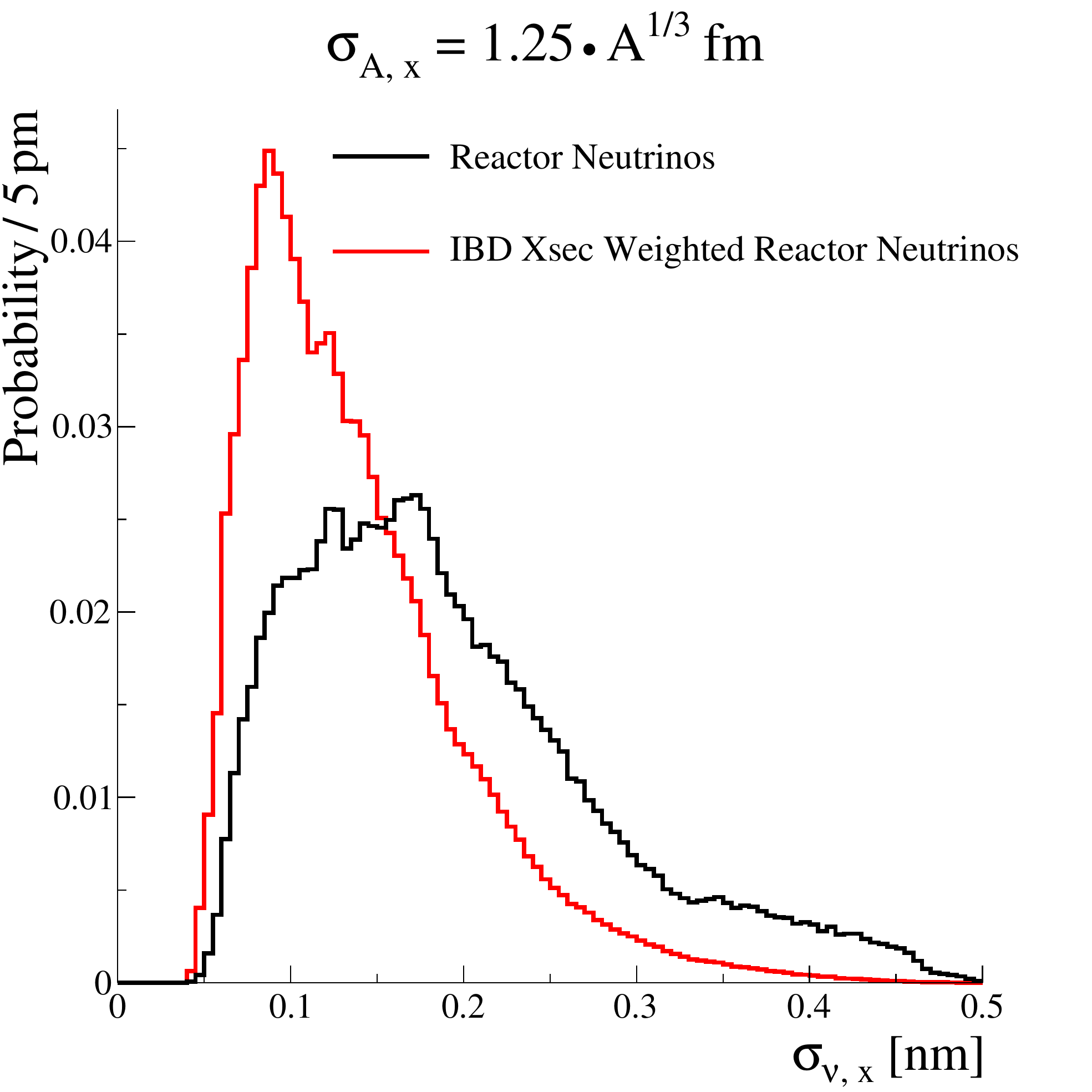}
\includegraphics[width=0.45\textwidth]{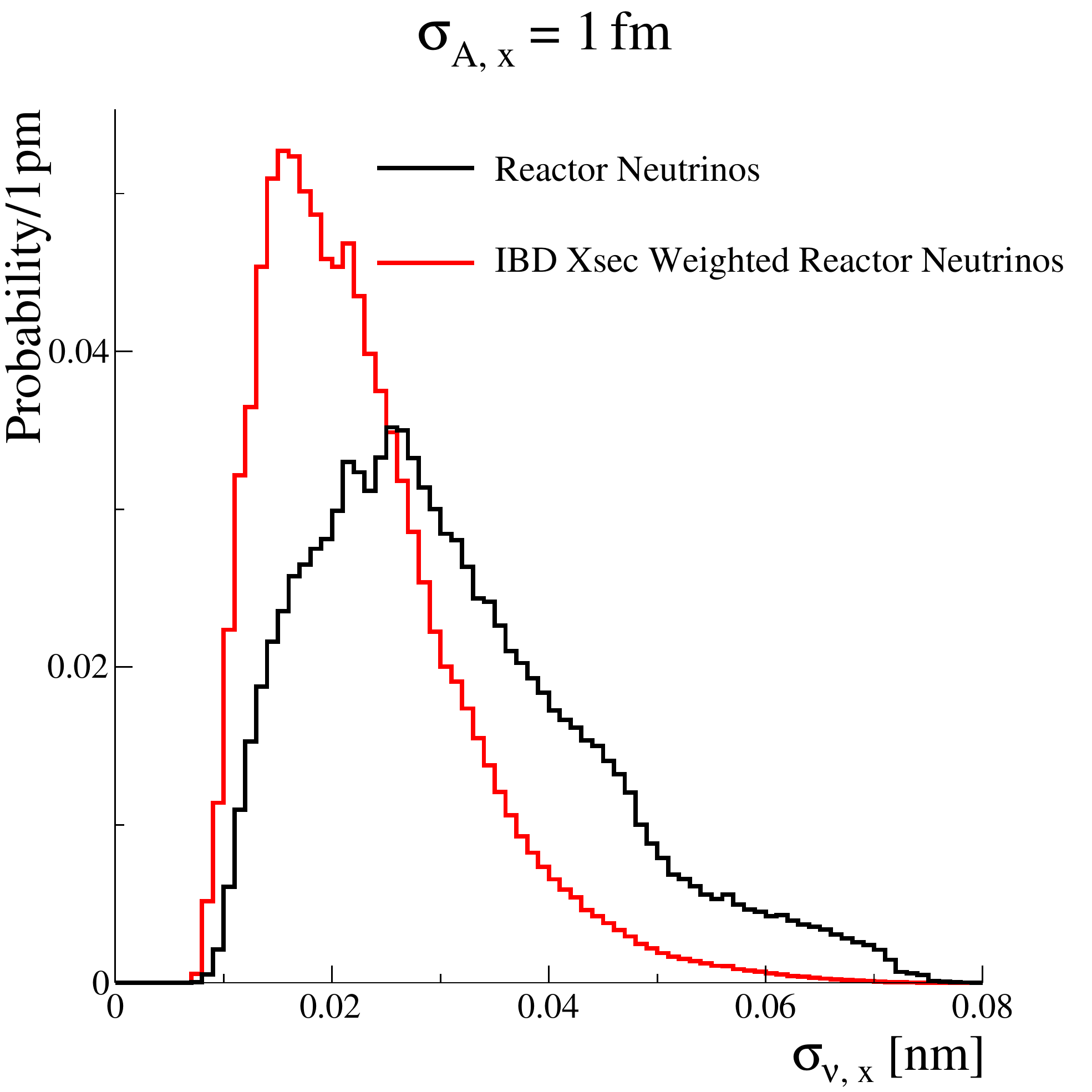}
\end{center}
\vspace{-.5cm}
\caption{The distributions of $\sigma_{\nu,x}$ values contributing to the electron antineutrino flux above the IBD threshold (black) and the IBD cross-section weighted flux (or, rate with no oscillations; red), assuming $\sigma_{A,x}=1.25\cdot\mathrm{A}^{1/3}~\mathrm{fm}$ (left) and $\sigma_{A,x}=1$~fm (right).}
\label{sigma_nu_1d}
\end{figure*}

\begin{figure*}[ht]
\begin{center}
\includegraphics[width=.95\textwidth]{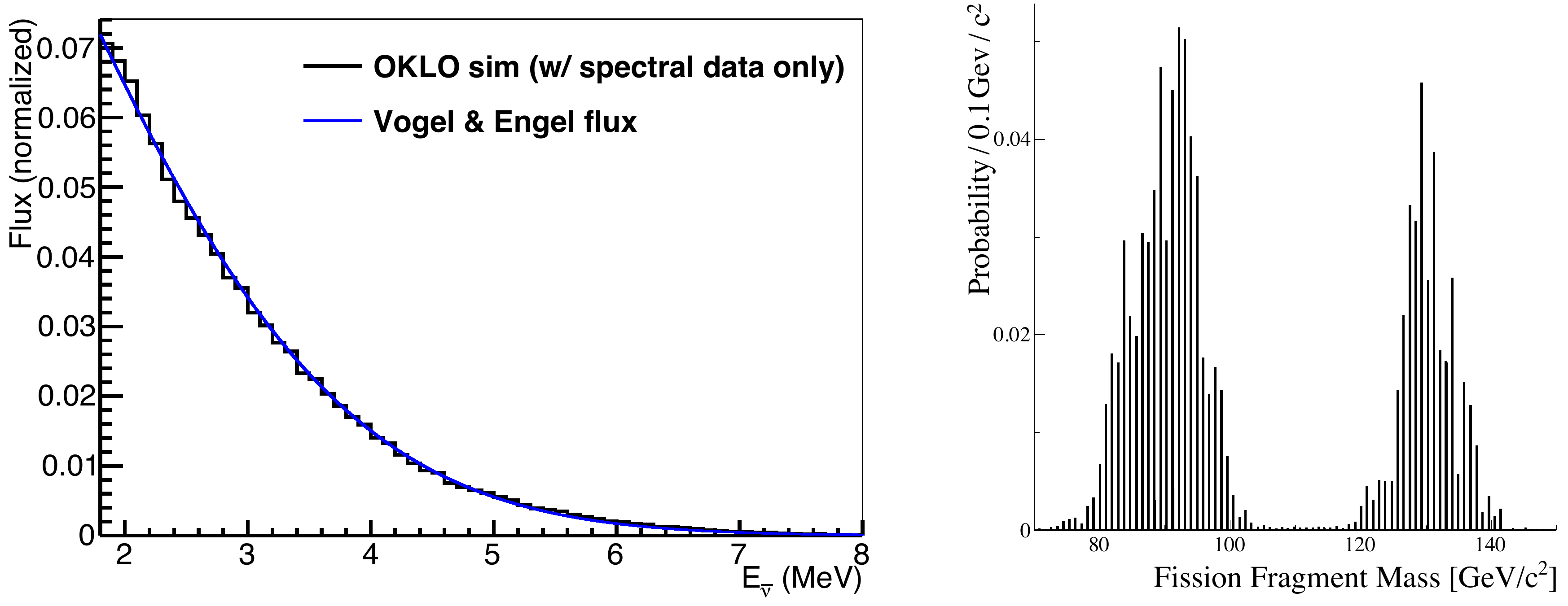}
\end{center}
\vspace{-.5cm}
\caption{(Left) The reactor electron antineutrino flux corresponding to the spectral data in OKLO alongside the Vogel \& Engel flux prediction~\cite{Vogel:1989iv}. (Right) The distribution of beta-decaying fission fragment masses that produce the electron antineutrino flux above the IBD threshold ($E_{\overline{\nu}_e}=1.8$~MeV).}
\label{flux_fission_frag}
\end{figure*}

The simulation uses the spectral data of 294~known fission daughters, including beta branches, representing about 35\% of the total known daughters in the simulation above the 1.8~MeV inverse beta decay (IBD; $\overline{\nu}_e p \rightarrow e^+ n$) interaction threshold. Spectral data information is available for an estimated $\sim$90\% of the total known daughter nuclide decay rate, per initial actinide fission, contributions above the IBD interaction threshold. We expect the quantitative conclusions of this article to be largely unaffected by the missing spectral data. With the OKLO information in hand, including relative normalizations amongst the parent nuclides and branches, insofar as they contribute to the antineutrino spectrum, parent/daughter masses, and $Q$-values, we then simulate individual beta decays according to the phase space available. Assuming that the parent width is equal to the nuclear diameter, approximated with $\sigma_{A,x}=1.25\cdot\mathrm{A}^{1/3}~\mathrm{fm}$, and that the antineutrino width is found using the parent width according to Eq.~\ref{sigma_nu_equation}, we arrive at the relationship between $\sigma_{\nu,x}$ and $E_{\overline{\nu}_e}$ shown in Fig.~\ref{sigmax_vs_energy_nuclear_width} (top). The analogous plot but assuming that the parent width is equal to the nucleon-nucleon correlation length scale, approximated with $\sigma_{A,x}=1~\mathrm{fm}$, is shown as well. The main figures show the values most relevant for reactor antineutrino experiments ($E_{\overline{\nu}_e}>1.8$~MeV), while the insets show lower energies. Figure~\ref{sigma_nu_1d} shows a projection of these plots in terms of the $\sigma_{\nu,x}$ values that contribute to the electron antineutrino flux and unoscillated rate above the IBD threshold. A projection of the plot(s) along the antineutrino energy axis is also shown in Fig.~\ref{flux_fission_frag} alongside the Vogel \& Engel prediction~\cite{Vogel:1989iv}. As can be seen, the antineutrino flux associated with the beta decay spectral data captures the overall flux well, despite the missing data. The fission fragments that contribute to the flux are also shown in the figure. 

Figure~\ref{sigmax_vs_energy_nuclear_width} shows that reactor antineutrinos will have a distribution of wavepacket widths that varies with antineutrino energy. Generalizing this prediction to cover both the nucleonic correlations and nuclear diameter distance scales discussed above in setting the parent width, we find $\sigma_{\nu,x}\sim10\mathrm{-}400$~pm for $E_{\overline{\nu}_e}>1.8$~MeV, which captures the full range of dependence on the parent properties, distance scales, and kinematics of the decay, including our oversimplified treatment of correlations among nucleons. Notably, the approximation $\sigma_{\nu,x} \approx 20000\cdot \sigma_{A,x}$  provides accurate results for most oscillation calculations.

As discussed in detail in Refs.~\cite{deGouvea:2020hfl,JUNO:2021ydg,Marzec:2022mcz}, the JUNO experiment provides a sensitive testing ground for the characteristic size of a reactor-induced antineutrino wavepacket based on its effect on observable electron antineutrino disappearance probability. The three-neutrino oscillation probability equation is modified to include a finite neutrino wavepacket via an expansion of Eq.~\ref{osc_eq}, $P_{ee} =1-P_{21}-P_{31}-P_{32}$ with

\begin{eqnarray}
P_{21}&& =\cos^4\theta_{13}\sin^2 2\theta_{12}\cdot \frac{1}{2}\Big[1-\cos\frac{1.27\Delta m^2_{21}L}{E}\cdot\nonumber\\&&\mathrm{exp}(-\frac{L^2(\Delta m^2_{21})^2}{32E^4\sigma^2_{\nu,x}})\Big]\nonumber \\
P_{31}&& =\cos^2\theta_{12}\sin^2 2\theta_{13}\cdot \frac{1}{2}\Big[1-\cos\frac{1.27\Delta m^2_{31}L}{E}\cdot\nonumber\\&&\mathrm{exp}(-\frac{L^2(\Delta m^2_{31})^2}{32E^4\sigma^2_{\nu,x}})\Big] \nonumber \\
P_{32}&& =\sin^2\theta_{12}\sin^2 2\theta_{13}\cdot \frac{1}{2}\Big[1-\cos\frac{1.27\Delta m^2_{32}L}{E}\cdot\nonumber\\&&\mathrm{exp}(-\frac{L^2(\Delta m^2_{32})^2}{32E^4\sigma^2_{\nu,x}})\Big] ~.
\label{Pee_sigmax}
\end{eqnarray} 

\begin{figure}[t]
\begin{centering}
\includegraphics[width=11cm]{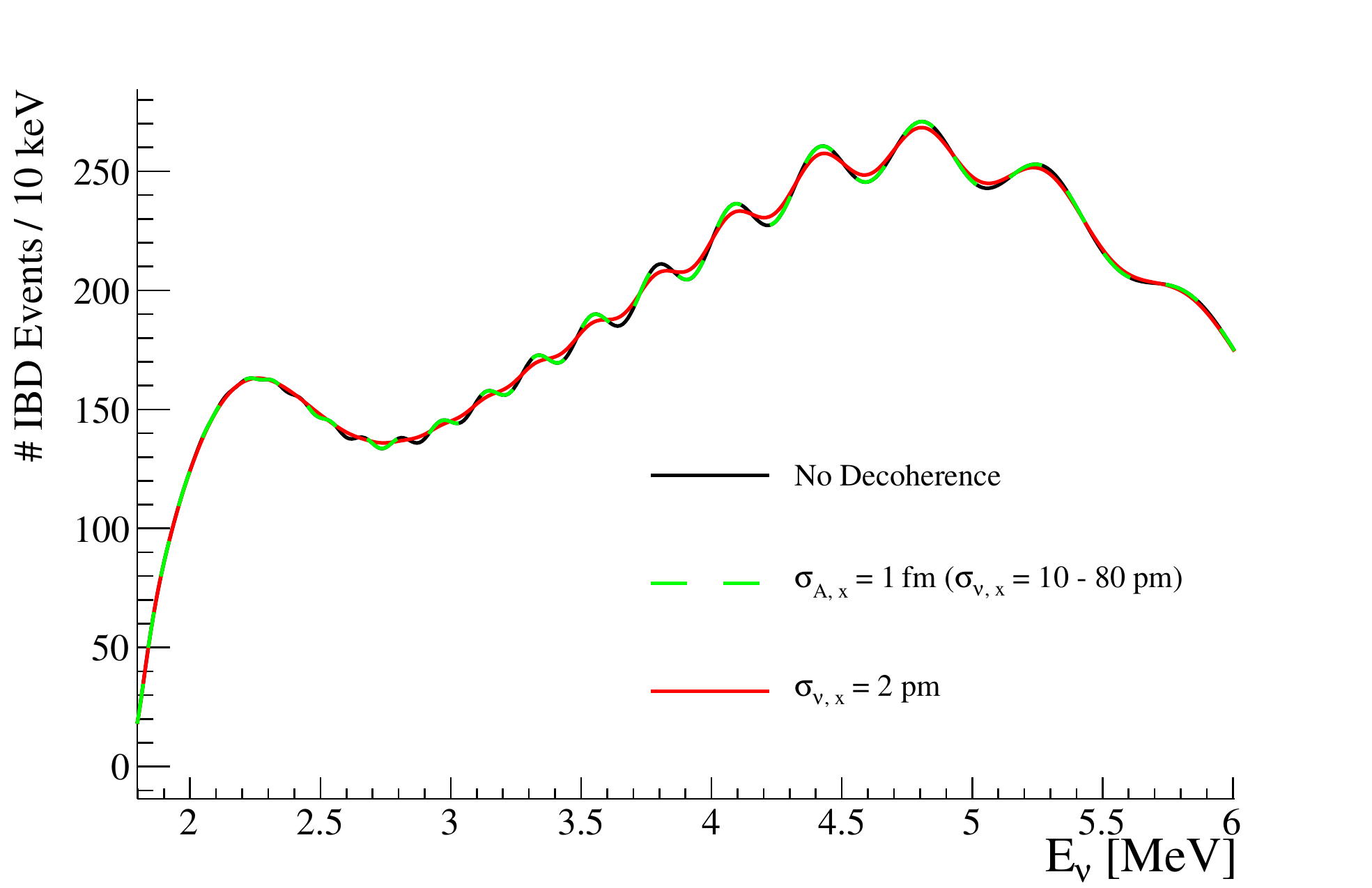}
\vspace{-.3cm}
\caption{The expected electron antineutrino energy spectrum in JUNO after 6~years of running, including energy resolution smearing effects, under a few wavepacket width assumptions.}
\label{juno_spectrum_example}
\end{centering}
\end{figure}

With this equation and the information in Fig.~\ref{sigmax_vs_energy_nuclear_width}, we can predict the impact on the observable electron antineutrino energy spectrum in JUNO. Figure~\ref{juno_spectrum_example} shows the event rate expectation ($10^5$ events total, after oscillations) using the JUNO experimental assumptions described in Refs.~\cite{Marzec:2022mcz,JUNO_2022_physics_paper} including, among others, 6~years of running, 
a 20~kton liquid scintillator far detector $\sim$53~km from a set of 26.6~GWth PWR reactor complexes, energy resolution effects, as well as oscillation parameters given by NuFIT\,5.0~\cite{Esteban:2020cvm}. As can be seen, JUNO is insensitive to even the lower end of our predicted range, $\sigma_{\nu,x}\sim10\mathrm{-}80$~pm (corresponding to $\sigma_{A,x}$=1~fm), which is consistent with no decoherence. Notably, however, the lowest end of this prediction is only about a factor of 3 higher than JUNO's expected sensitivity (see Fig.~\ref{juno_spectrum_example} for the observable oscillation effect of $\sigma_{\nu,x}=2$~pm, as an example)--but, there is no realistic way to bridge the gap without invoking a longer baseline \textit{and} bigger detector option. Beyond JUNO, there may be a possibility to observe the wavepacket effect with a future experiment in the case that a high-$\Delta m^2$ mass splitting ($\Delta m^2\gtrsim10~\mathrm{eV}^2$) is participating in oscillations, but we leave this study for future work.

\section{Conclusions~\label{Conclusions}}

In this paper we have used the basic principles of quantum mechanics of entangled systems to predict the wavepacket widths for antineutrinos created via beta decaying fission fragments that produce the antineutrino flux from nuclear reactors.  Accounting for the hierarchy of entanglement scales of the decaying parent we find that the antineutrino width is dictated by the scale with which nucleons inside the nucleus encode each others' location, even if the nucleus in question is quantum mechanically delocalized.  This scale is determined by both the size of the nucleus and the distribution of nucleon-nucleon correlation ranges within it.  Using the resulting density matrix as input to an oscillation calculation that properly accounts for the emission of the entangled recoil, we predict the distance and energy scales of non-trivial coherence loss effects from wavepacket separation with non-trivial dependencies on antineutrino energy, decay parent, and recoil system kinematics.  It is notable that because each neutrino in three-body decays is entangled with a different recoiling final state, the coherent quantum mechanical width of the neutrino that determines oscillation coherence (the off-diagonal width of the density matrix) is in general far narrower than the total width of the energy spectrum (the diagonal width of the neutrino density matrix).  

While our conclusion is similar, our prediction of the wavepacket widths is not in quantitative agreement with some others that have been given in recent literature.  For example, Ref.~\cite{akhmedov2022damping} posits that the distance scale between scatters of the recoil system provides the limiting distance scale for neutrino coherence.  We have shown in Sec.~\ref{sec:Finding-the-delocalization} that even with a highly delocalized nucleus, the sub-nuclear distance scale effects remain limiting when all factors are properly accounted for in the final state. We also consider definitive the argument that the final state entangled recoil cannot influence the coherence behaviour of the neutrino, for reasons previously outlined in the Appendix of Ref.~\cite{jones2015dynamical} and elaborated in Ref.~\cite{jones2022comment}.  The treatment used in this paper derives the coherence loss effect directly from the principles of entangled-system quantum mechanics, resulting in a clear statement about the expected coherence distances with little room for arbitrariness. One notable deficiency in our treatment, however, is that we have used only a representative example for the nuclear wave function, since a direct calculation remains elusive for the large fission fragments that emit antineutrinos in nuclear reactors. Incorporation of an \textit{ab-initio} wave function for a small nucleus will be the subject of a forthcoming study.

Unfortunately, the coherence loss effects that we predict are outside of current and foreseeable experimental sensitivity, including at the JUNO reactor antineutrino experiment.  Still, observation of the coherence loss effect on oscillations is something to strive for, as it would both provide the first observation of a novel and to-date unmeasured neutrino property: the quantum mechanical localization scale of an antineutrino from beta decay at emission. Such an observation would also represent a fundamental test of the validity of quantum mechanics. The wavepacket width is dictated by the scale to which the environment around the open quantum system of parent-neutrino-recoil system encodes information about it through entanglement. This principle, when combined with a quantum measurement process yielding definite outcomes given a probabilistically mixed state, is responsible for the emergence of classical behaviour in quantum systems in general.  Its accurate prediction and subsequent measurement thus represents a test of the theory of the emergence of classicality. That such a principle can be tested, even in the far future, through accurate antineutrino interferomery would represent another entry in the list of opportunities neutrinos have afforded us as experimenters, to better understand the Universe and our role within it.

\section*{Note added: The choice of nuclear wave function}

V1 of this paper featured a slightly different example nuclear wave function Eq.~\ref{eq:COMWF} that omitted the center-of-mass forcing delta function.  Examination of that scenario led to the quantitatively identical results in all relevant conditions, but subtle differences when less realistic parameter choices are made.  Our position is that either Eq.~\ref{eq:COMWF} with or without the delta function would be reasonable choices for investigating the relevant distance scales of quantum coherence.

The reason we have opted to include the delta function in Eq.~\ref{eq:COMWF} is that it ensures that the function $\phi(X)$ of Eq.~\ref{eq:PhiDef} is equivalent to a wave function of the center of mass coordinate $X$. Absent this delta function, $\phi(X)$ would represent a smearing function associated with delocalization in the material, but not a center of mass wave function in a strictest sense. This point was brought to our attention as part of fruitful discussions about our paper with Evgeny Akhmedov and Alexei Smirnov.

One motivation for favoring a model wave function separable in this manner is that such a factorization emerges when solving the unitary time evolution of systems whose dynamics are governed by potentials of the form $V(X)+V(x_a-x_b)$.  We note that these are not strictly accurate assumptions for present purposes. In this work  $\Psi$ is the result of time-evolving a larger quantum system and then tracing out environmental degrees of freedom to obtain a mixed state, followed by decomposing that mixed state into pure-state contributions per Eq.~\ref{eq:RhoDecomp}.  As such, either Eq.~\ref{eq:COMWF} with or without a delta function seem to  provide suitably illustrative model systems. 

The difference between inclusion (in the current manuscript) vs non-inclusion (in V1 of the manuscript)  of a delta function in Eq.~\ref{eq:COMWF} is manifest only in the limit $\sigma_M\ll \sigma_N,\sigma_c$.  In the latter case the nuclear size itself becomes the limiting scale, in contrast to the present case where the localization scale tends to zero along with $\sigma_M$. This difference is unsurprising, since the physical interpretation of $\sigma_M$ for the two systems is different in this limit. In the physically relevant cases $\sigma_N,\sigma_C\geq\sigma_M$, both quantitative and qualitative results are equivalent between the two cases.

\section*{Acknowledgements}

We thank Saori Pastore, Raquel Castillo Fernandez, Carlos Arguelles, Krishan Mistry, Jonathan Asaadi, and Bryce Littlejohn for helpful discussions on the subject of this manuscript. We are also grateful to Evgeny Akhmedov and Alexei Smirnov for fruitful discussions and critical remarks. BJPJ is supported by Department of Energy under Award numbers {DE-SC0019054} and {DE-SC0019223} and the National Science Foundation via the IceCube South Pole Neutrino Observatory. EM and JS are supported by the Department of Energy, Office of Science, under Award No. DE-SC0007859 and the Heising-Simons Foundation.

\bibliography{main}

\begin{thebibliography}{42}%
\makeatletter
\providecommand \@ifxundefined [1]{%
 \@ifx{#1\undefined}
}%
\providecommand \@ifnum [1]{%
 \ifnum #1\expandafter \@firstoftwo
 \else \expandafter \@secondoftwo
 \fi
}%
\providecommand \@ifx [1]{%
 \ifx #1\expandafter \@firstoftwo
 \else \expandafter \@secondoftwo
 \fi
}%
\providecommand \natexlab [1]{#1}%
\providecommand \enquote  [1]{``#1''}%
\providecommand \bibnamefont  [1]{#1}%
\providecommand \bibfnamefont [1]{#1}%
\providecommand \citenamefont [1]{#1}%
\providecommand \href@noop [0]{\@secondoftwo}%
\providecommand \href [0]{\begingroup \@sanitize@url \@href}%
\providecommand \@href[1]{\@@startlink{#1}\@@href}%
\providecommand \@@href[1]{\endgroup#1\@@endlink}%
\providecommand \@sanitize@url [0]{\catcode `\\12\catcode `\$12\catcode
  `\&12\catcode `\#12\catcode `\^12\catcode `\_12\catcode `\%12\relax}%
\providecommand \@@startlink[1]{}%
\providecommand \@@endlink[0]{}%
\providecommand \url  [0]{\begingroup\@sanitize@url \@url }%
\providecommand \@url [1]{\endgroup\@href {#1}{\urlprefix }}%
\providecommand \urlprefix  [0]{URL }%
\providecommand \Eprint [0]{\href }%
\providecommand \doibase [0]{http://dx.doi.org/}%
\providecommand \selectlanguage [0]{\@gobble}%
\providecommand \bibinfo  [0]{\@secondoftwo}%
\providecommand \bibfield  [0]{\@secondoftwo}%
\providecommand \translation [1]{[#1]}%
\providecommand \BibitemOpen [0]{}%
\providecommand \bibitemStop [0]{}%
\providecommand \bibitemNoStop [0]{.\EOS\space}%
\providecommand \EOS [0]{\spacefactor3000\relax}%
\providecommand \BibitemShut  [1]{\csname bibitem#1\endcsname}%
\let\auto@bib@innerbib\@empty
\bibitem [{\citenamefont {Kayser}\ and\ \citenamefont
  {Kopp}(2010)}]{kayser2010testing}%
  \BibitemOpen
  \bibfield  {author} {\bibinfo {author} {\bibfnamefont {B.}~\bibnamefont
  {Kayser}}\ and\ \bibinfo {author} {\bibfnamefont {J.}~\bibnamefont {Kopp}},\
  }\href@noop {} {\  (\bibinfo {year} {2010})},\ \Eprint
  {http://arxiv.org/abs/1005.4081} {arXiv:1005.4081 [hep-ph]} \BibitemShut
  {NoStop}%
\bibitem [{\citenamefont {Akhmedov}\ \emph {et~al.}(2012)\citenamefont
  {Akhmedov}, \citenamefont {Hernandez},\ and\ \citenamefont
  {Smirnov}}]{akhmedov2012neutrino}%
  \BibitemOpen
  \bibfield  {author} {\bibinfo {author} {\bibfnamefont {E.}~\bibnamefont
  {Akhmedov}}, \bibinfo {author} {\bibfnamefont {D.}~\bibnamefont {Hernandez}},
  \ and\ \bibinfo {author} {\bibfnamefont {A.}~\bibnamefont {Smirnov}},\ }\href
  {\doibase 10.1007/JHEP04(2012)052} {\bibfield  {journal} {\bibinfo  {journal}
  {JHEP}\ }\textbf {\bibinfo {volume} {04}},\ \bibinfo {pages} {052} (\bibinfo
  {year} {2012})},\ \Eprint {http://arxiv.org/abs/1201.4128} {arXiv:1201.4128
  [hep-ph]} \BibitemShut {NoStop}%
\bibitem [{\citenamefont {Jones}(2015)}]{jones2015dynamical}%
  \BibitemOpen
  \bibfield  {author} {\bibinfo {author} {\bibfnamefont {B.~J.~P.}\
  \bibnamefont {Jones}},\ }\href {\doibase 10.1103/PhysRevD.91.053002}
  {\bibfield  {journal} {\bibinfo  {journal} {Phys. Rev. D}\ }\textbf {\bibinfo
  {volume} {91}},\ \bibinfo {pages} {053002} (\bibinfo {year} {2015})},\
  \Eprint {http://arxiv.org/abs/1412.2264} {arXiv:1412.2264 [hep-ph]}
  \BibitemShut {NoStop}%
\bibitem [{\citenamefont {Coloma}\ \emph {et~al.}(2018)\citenamefont {Coloma},
  \citenamefont {Lopez-Pavon}, \citenamefont {Martinez-Soler},\ and\
  \citenamefont {Nunokawa}}]{coloma2018decoherence}%
  \BibitemOpen
  \bibfield  {author} {\bibinfo {author} {\bibfnamefont {P.}~\bibnamefont
  {Coloma}}, \bibinfo {author} {\bibfnamefont {J.}~\bibnamefont {Lopez-Pavon}},
  \bibinfo {author} {\bibfnamefont {I.}~\bibnamefont {Martinez-Soler}}, \ and\
  \bibinfo {author} {\bibfnamefont {H.}~\bibnamefont {Nunokawa}},\ }\href
  {\doibase 10.1140/epjc/s10052-018-6092-6} {\bibfield  {journal} {\bibinfo
  {journal} {Eur. Phys. J. C}\ }\textbf {\bibinfo {volume} {78}},\ \bibinfo
  {pages} {614} (\bibinfo {year} {2018})},\ \Eprint
  {http://arxiv.org/abs/1803.04438} {arXiv:1803.04438 [hep-ph]} \BibitemShut
  {NoStop}%
\bibitem [{\citenamefont {Coelho}\ \emph {et~al.}(2017)\citenamefont {Coelho},
  \citenamefont {Mann},\ and\ \citenamefont {Bashar}}]{coelho2017nonmaximal}%
  \BibitemOpen
  \bibfield  {author} {\bibinfo {author} {\bibfnamefont {J.~A.~B.}\
  \bibnamefont {Coelho}}, \bibinfo {author} {\bibfnamefont {W.~A.}\
  \bibnamefont {Mann}}, \ and\ \bibinfo {author} {\bibfnamefont {S.~S.}\
  \bibnamefont {Bashar}},\ }\href {\doibase 10.1103/PhysRevLett.118.221801}
  {\bibfield  {journal} {\bibinfo  {journal} {Phys. Rev. Lett.}\ }\textbf
  {\bibinfo {volume} {118}},\ \bibinfo {pages} {221801} (\bibinfo {year}
  {2017})}\BibitemShut {NoStop}%
\bibitem [{\citenamefont {Gomes}\ \emph {et~al.}(2019)\citenamefont {Gomes},
  \citenamefont {Forero}, \citenamefont {Guzzo}, \citenamefont {de~Holanda},\
  and\ \citenamefont {Oliveira}}]{gomes2019quantum}%
  \BibitemOpen
  \bibfield  {author} {\bibinfo {author} {\bibfnamefont {G.~B.}\ \bibnamefont
  {Gomes}}, \bibinfo {author} {\bibfnamefont {D.~V.}\ \bibnamefont {Forero}},
  \bibinfo {author} {\bibfnamefont {M.~M.}\ \bibnamefont {Guzzo}}, \bibinfo
  {author} {\bibfnamefont {P.~C.}\ \bibnamefont {de~Holanda}}, \ and\ \bibinfo
  {author} {\bibfnamefont {R.~L.~N.}\ \bibnamefont {Oliveira}},\ }\href
  {\doibase 10.1103/PhysRevD.100.055023} {\bibfield  {journal} {\bibinfo
  {journal} {Phys. Rev. D}\ }\textbf {\bibinfo {volume} {100}},\ \bibinfo
  {pages} {055023} (\bibinfo {year} {2019})}\BibitemShut {NoStop}%
\bibitem [{\citenamefont {Akhmedov}\ and\ \citenamefont
  {Smirnov}(2009)}]{akhmedov2009paradoxes}%
  \BibitemOpen
  \bibfield  {author} {\bibinfo {author} {\bibfnamefont {E.~K.}\ \bibnamefont
  {Akhmedov}}\ and\ \bibinfo {author} {\bibfnamefont {A.~Y.}\ \bibnamefont
  {Smirnov}},\ }\href {\doibase 10.1134/S1063778809080122} {\bibfield
  {journal} {\bibinfo  {journal} {Phys. Atom. Nucl.}\ }\textbf {\bibinfo
  {volume} {72}},\ \bibinfo {pages} {1363} (\bibinfo {year} {2009})},\ \Eprint
  {http://arxiv.org/abs/0905.1903} {arXiv:0905.1903 [hep-ph]} \BibitemShut
  {NoStop}%
\bibitem [{\citenamefont {Giunti}\ and\ \citenamefont
  {Kim}(1998)}]{giunti1998coherence}%
  \BibitemOpen
  \bibfield  {author} {\bibinfo {author} {\bibfnamefont {C.}~\bibnamefont
  {Giunti}}\ and\ \bibinfo {author} {\bibfnamefont {C.~W.}\ \bibnamefont
  {Kim}},\ }\href {\doibase 10.1103/PhysRevD.58.017301} {\bibfield  {journal}
  {\bibinfo  {journal} {Phys. Rev. D}\ }\textbf {\bibinfo {volume} {58}},\
  \bibinfo {pages} {017301} (\bibinfo {year} {1998})},\ \Eprint
  {http://arxiv.org/abs/hep-ph/9711363} {arXiv:hep-ph/9711363} \BibitemShut
  {NoStop}%
\bibitem [{\citenamefont {Giunti}(2002)}]{giunti2002neutrino}%
  \BibitemOpen
  \bibfield  {author} {\bibinfo {author} {\bibfnamefont {C.}~\bibnamefont
  {Giunti}},\ }\href {\doibase 10.1088/1126-6708/2002/11/017} {\bibfield
  {journal} {\bibinfo  {journal} {JHEP}\ }\textbf {\bibinfo {volume} {11}},\
  \bibinfo {pages} {017} (\bibinfo {year} {2002})},\ \Eprint
  {http://arxiv.org/abs/hep-ph/0205014} {arXiv:hep-ph/0205014} \BibitemShut
  {NoStop}%
\bibitem [{\citenamefont {Cohen}\ \emph {et~al.}(2009)\citenamefont {Cohen},
  \citenamefont {Glashow},\ and\ \citenamefont
  {Ligeti}}]{cohen2009disentangling}%
  \BibitemOpen
  \bibfield  {author} {\bibinfo {author} {\bibfnamefont {A.~G.}\ \bibnamefont
  {Cohen}}, \bibinfo {author} {\bibfnamefont {S.~L.}\ \bibnamefont {Glashow}},
  \ and\ \bibinfo {author} {\bibfnamefont {Z.}~\bibnamefont {Ligeti}},\ }\href
  {\doibase 10.1016/j.physletb.2009.06.020} {\bibfield  {journal} {\bibinfo
  {journal} {Phys. Lett. B}\ }\textbf {\bibinfo {volume} {678}},\ \bibinfo
  {pages} {191} (\bibinfo {year} {2009})},\ \Eprint
  {http://arxiv.org/abs/0810.4602} {arXiv:0810.4602 [hep-ph]} \BibitemShut
  {NoStop}%
\bibitem [{\citenamefont {Wu}\ \emph {et~al.}(2010)\citenamefont {Wu},
  \citenamefont {Hutasoit}, \citenamefont {Boyanovsky},\ and\ \citenamefont
  {Holman}}]{wu2010dynamics}%
  \BibitemOpen
  \bibfield  {author} {\bibinfo {author} {\bibfnamefont {J.}~\bibnamefont
  {Wu}}, \bibinfo {author} {\bibfnamefont {J.~A.}\ \bibnamefont {Hutasoit}},
  \bibinfo {author} {\bibfnamefont {D.}~\bibnamefont {Boyanovsky}}, \ and\
  \bibinfo {author} {\bibfnamefont {R.}~\bibnamefont {Holman}},\ }\href
  {\doibase 10.1103/PhysRevD.82.013006} {\bibfield  {journal} {\bibinfo
  {journal} {Phys. Rev. D}\ }\textbf {\bibinfo {volume} {82}},\ \bibinfo
  {pages} {013006} (\bibinfo {year} {2010})},\ \Eprint
  {http://arxiv.org/abs/1005.3260} {arXiv:1005.3260 [hep-ph]} \BibitemShut
  {NoStop}%
\bibitem [{\citenamefont {Wu}\ \emph {et~al.}(2011)\citenamefont {Wu},
  \citenamefont {Hutasoit}, \citenamefont {Boyanovsky},\ and\ \citenamefont
  {Holman}}]{wu2011neutrino}%
  \BibitemOpen
  \bibfield  {author} {\bibinfo {author} {\bibfnamefont {J.}~\bibnamefont
  {Wu}}, \bibinfo {author} {\bibfnamefont {J.~A.}\ \bibnamefont {Hutasoit}},
  \bibinfo {author} {\bibfnamefont {D.}~\bibnamefont {Boyanovsky}}, \ and\
  \bibinfo {author} {\bibfnamefont {R.}~\bibnamefont {Holman}},\ }\href
  {\doibase 10.1142/S0217751X11054954} {\bibfield  {journal} {\bibinfo
  {journal} {Int. J. Mod. Phys. A}\ }\textbf {\bibinfo {volume} {26}},\
  \bibinfo {pages} {5261} (\bibinfo {year} {2011})},\ \Eprint
  {http://arxiv.org/abs/1002.2649} {arXiv:1002.2649 [hep-ph]} \BibitemShut
  {NoStop}%
\bibitem [{\citenamefont {An}\ \emph {et~al.}(2016)\citenamefont {An} \emph
  {et~al.}}]{JUNO:2015zny}%
  \BibitemOpen
  \bibfield  {author} {\bibinfo {author} {\bibfnamefont {F.}~\bibnamefont {An}}
  \emph {et~al.} (\bibinfo {collaboration} {JUNO}),\ }\href {\doibase
  10.1088/0954-3899/43/3/030401} {\bibfield  {journal} {\bibinfo  {journal} {J.
  Phys. G}\ }\textbf {\bibinfo {volume} {43}},\ \bibinfo {pages} {030401}
  (\bibinfo {year} {2016})},\ \Eprint {http://arxiv.org/abs/1507.05613}
  {arXiv:1507.05613 [physics.ins-det]} \BibitemShut {NoStop}%
\bibitem [{\citenamefont {Abusleme}\ \emph {et~al.}(2022)\citenamefont
  {Abusleme} \emph {et~al.}}]{JUNO_2022_physics_paper}%
  \BibitemOpen
  \bibfield  {author} {\bibinfo {author} {\bibfnamefont {A.}~\bibnamefont
  {Abusleme}} \emph {et~al.} (\bibinfo {collaboration} {JUNO}),\ }\href@noop {}
  {\  (\bibinfo {year} {2022})},\ \Eprint {http://arxiv.org/abs/2204.13249}
  {arXiv:2204.13249 [hep-ex]} \BibitemShut {NoStop}%
\bibitem [{\citenamefont {de~Gouv\^ea}\ \emph {et~al.}(2020)\citenamefont
  {de~Gouv\^ea}, \citenamefont {de~Romeri},\ and\ \citenamefont
  {Ternes}}]{deGouvea:2020hfl}%
  \BibitemOpen
  \bibfield  {author} {\bibinfo {author} {\bibfnamefont {A.}~\bibnamefont
  {de~Gouv\^ea}}, \bibinfo {author} {\bibfnamefont {V.}~\bibnamefont
  {de~Romeri}}, \ and\ \bibinfo {author} {\bibfnamefont {C.~A.}\ \bibnamefont
  {Ternes}},\ }\href {\doibase 10.1007/JHEP08(2020)049} {\bibfield  {journal}
  {\bibinfo  {journal} {JHEP}\ }\textbf {\bibinfo {volume} {08}},\ \bibinfo
  {pages} {018} (\bibinfo {year} {2020})},\ \Eprint
  {http://arxiv.org/abs/2005.03022} {arXiv:2005.03022 [hep-ph]} \BibitemShut
  {NoStop}%
\bibitem [{\citenamefont {Wang}\ \emph {et~al.}(2022)\citenamefont {Wang} \emph
  {et~al.}}]{JUNO:2021ydg}%
  \BibitemOpen
  \bibfield  {author} {\bibinfo {author} {\bibfnamefont {J.}~\bibnamefont
  {Wang}} \emph {et~al.} (\bibinfo {collaboration} {JUNO}),\ }\href {\doibase
  10.1007/JHEP06(2022)062} {\bibfield  {journal} {\bibinfo  {journal} {JHEP}\
  }\textbf {\bibinfo {volume} {06}},\ \bibinfo {pages} {062} (\bibinfo {year}
  {2022})},\ \Eprint {http://arxiv.org/abs/2112.14450} {arXiv:2112.14450
  [hep-ex]} \BibitemShut {NoStop}%
\bibitem [{\citenamefont {Marzec}\ and\ \citenamefont
  {Spitz}(2022)}]{Marzec:2022mcz}%
  \BibitemOpen
  \bibfield  {author} {\bibinfo {author} {\bibfnamefont {E.}~\bibnamefont
  {Marzec}}\ and\ \bibinfo {author} {\bibfnamefont {J.}~\bibnamefont {Spitz}},\
  }\href {\doibase 10.1103/PhysRevD.106.053007} {\bibfield  {journal} {\bibinfo
   {journal} {Phys. Rev. D}\ }\textbf {\bibinfo {volume} {106}},\ \bibinfo
  {pages} {053007} (\bibinfo {year} {2022})},\ \Eprint
  {http://arxiv.org/abs/2208.04277} {arXiv:2208.04277 [hep-ph]} \BibitemShut
  {NoStop}%
\bibitem [{\citenamefont {de~Gouv\^ea}\ \emph {et~al.}(2021)\citenamefont
  {de~Gouv\^ea}, \citenamefont {De~Romeri},\ and\ \citenamefont
  {Ternes}}]{deGouvea:2021uvg}%
  \BibitemOpen
  \bibfield  {author} {\bibinfo {author} {\bibfnamefont {A.}~\bibnamefont
  {de~Gouv\^ea}}, \bibinfo {author} {\bibfnamefont {V.}~\bibnamefont
  {De~Romeri}}, \ and\ \bibinfo {author} {\bibfnamefont {C.~A.}\ \bibnamefont
  {Ternes}},\ }\href {\doibase 10.1007/JHEP06(2021)042} {\bibfield  {journal}
  {\bibinfo  {journal} {JHEP}\ }\textbf {\bibinfo {volume} {06}},\ \bibinfo
  {pages} {042} (\bibinfo {year} {2021})},\ \Eprint
  {http://arxiv.org/abs/2104.05806} {arXiv:2104.05806 [hep-ph]} \BibitemShut
  {NoStop}%
\bibitem [{\citenamefont {An}\ \emph {et~al.}(2017)\citenamefont {An} \emph
  {et~al.}}]{DayaBay:2016ouy}%
  \BibitemOpen
  \bibfield  {author} {\bibinfo {author} {\bibfnamefont {F.~P.}\ \bibnamefont
  {An}} \emph {et~al.} (\bibinfo {collaboration} {Daya Bay}),\ }\href {\doibase
  10.1140/epjc/s10052-017-4970-y} {\bibfield  {journal} {\bibinfo  {journal}
  {Eur. Phys. J. C}\ }\textbf {\bibinfo {volume} {77}},\ \bibinfo {pages} {606}
  (\bibinfo {year} {2017})},\ \Eprint {http://arxiv.org/abs/1608.01661}
  {arXiv:1608.01661 [hep-ex]} \BibitemShut {NoStop}%
\bibitem [{\citenamefont {Aguilar-Arevalo}\ \emph {et~al.}(2021)\citenamefont
  {Aguilar-Arevalo} \emph {et~al.}}]{miniboone_new}%
  \BibitemOpen
  \bibfield  {author} {\bibinfo {author} {\bibfnamefont {A.~A.}\ \bibnamefont
  {Aguilar-Arevalo}} \emph {et~al.} (\bibinfo {collaboration} {MiniBooNE
  Collaboration}),\ }\href {\doibase 10.1103/PhysRevD.103.052002} {\bibfield
  {journal} {\bibinfo  {journal} {Phys. Rev. D}\ }\textbf {\bibinfo {volume}
  {103}},\ \bibinfo {pages} {052002} (\bibinfo {year} {2021})}\BibitemShut
  {NoStop}%
\bibitem [{\citenamefont {Athanassopoulos}\ \emph {et~al.}(1998)\citenamefont
  {Athanassopoulos} \emph {et~al.}}]{lsnd3}%
  \BibitemOpen
  \bibfield  {author} {\bibinfo {author} {\bibfnamefont {C.}~\bibnamefont
  {Athanassopoulos}} \emph {et~al.} (\bibinfo {collaboration} {LSND
  Collaboration}),\ }\href {\doibase 10.1103/PhysRevLett.81.1774} {\bibfield
  {journal} {\bibinfo  {journal} {Phys. Rev. Lett.}\ }\textbf {\bibinfo
  {volume} {81}},\ \bibinfo {pages} {1774} (\bibinfo {year}
  {1998})}\BibitemShut {NoStop}%
\bibitem [{\citenamefont {Giunti}\ and\ \citenamefont
  {Laveder}(2011)}]{source}%
  \BibitemOpen
  \bibfield  {author} {\bibinfo {author} {\bibfnamefont {C.}~\bibnamefont
  {Giunti}}\ and\ \bibinfo {author} {\bibfnamefont {M.}~\bibnamefont
  {Laveder}},\ }\href {\doibase 10.1103/PhysRevC.83.065504} {\bibfield
  {journal} {\bibinfo  {journal} {Phys. Rev. C}\ }\textbf {\bibinfo {volume}
  {83}},\ \bibinfo {pages} {065504} (\bibinfo {year} {2011})}\BibitemShut
  {NoStop}%
\bibitem [{\citenamefont {Barinov}\ \emph {et~al.}(2022)\citenamefont {Barinov}
  \emph {et~al.}}]{Barinov:2021asz}%
  \BibitemOpen
  \bibfield  {author} {\bibinfo {author} {\bibfnamefont {V.~V.}\ \bibnamefont
  {Barinov}} \emph {et~al.},\ }\href {\doibase 10.1103/PhysRevLett.128.232501}
  {\bibfield  {journal} {\bibinfo  {journal} {Phys. Rev. Lett.}\ }\textbf
  {\bibinfo {volume} {128}},\ \bibinfo {pages} {232501} (\bibinfo {year}
  {2022})},\ \Eprint {http://arxiv.org/abs/2109.11482} {arXiv:2109.11482
  [nucl-ex]} \BibitemShut {NoStop}%
\bibitem [{\citenamefont {Arg\"uelles}\ \emph {et~al.}(2022)\citenamefont
  {Arg\"uelles}, \citenamefont {Bert\'olez-Mart\'\i{}nez},\ and\ \citenamefont
  {Salvado}}]{Arguelles:2022bvt}%
  \BibitemOpen
  \bibfield  {author} {\bibinfo {author} {\bibfnamefont {C.~A.}\ \bibnamefont
  {Arg\"uelles}}, \bibinfo {author} {\bibfnamefont {T.}~\bibnamefont
  {Bert\'olez-Mart\'\i{}nez}}, \ and\ \bibinfo {author} {\bibfnamefont
  {J.}~\bibnamefont {Salvado}},\ }\href@noop {} {\  (\bibinfo {year} {2022})},\
  \Eprint {http://arxiv.org/abs/2201.05108} {arXiv:2201.05108 [hep-ph]}
  \BibitemShut {NoStop}%
\bibitem [{\citenamefont {Zurek}(2003)}]{zurek2003decoherence}%
  \BibitemOpen
  \bibfield  {author} {\bibinfo {author} {\bibfnamefont {W.~H.}\ \bibnamefont
  {Zurek}},\ }\href {\doibase 10.1103/RevModPhys.75.715} {\bibfield  {journal}
  {\bibinfo  {journal} {Rev. Mod. Phys.}\ }\textbf {\bibinfo {volume} {75}},\
  \bibinfo {pages} {715} (\bibinfo {year} {2003})},\ \Eprint
  {http://arxiv.org/abs/quant-ph/0105127} {arXiv:quant-ph/0105127} \BibitemShut
  {NoStop}%
\bibitem [{\citenamefont {Joos}\ \emph {et~al.}(2003)\citenamefont {Joos},
  \citenamefont {Zeh}, \citenamefont {Giulini}, \citenamefont {Kiefer},
  \citenamefont {Kupsch},\ and\ \citenamefont
  {Stamatescu}}]{joos2013decoherence}%
  \BibitemOpen
  \bibfield  {author} {\bibinfo {author} {\bibfnamefont {E.}~\bibnamefont
  {Joos}}, \bibinfo {author} {\bibfnamefont {H.}~\bibnamefont {Zeh}}, \bibinfo
  {author} {\bibfnamefont {D.}~\bibnamefont {Giulini}}, \bibinfo {author}
  {\bibfnamefont {C.}~\bibnamefont {Kiefer}}, \bibinfo {author} {\bibfnamefont
  {J.}~\bibnamefont {Kupsch}}, \ and\ \bibinfo {author} {\bibfnamefont
  {I.}~\bibnamefont {Stamatescu}},\ }\href
  {https://books.google.com/books?id=6eTHcxeNxdUC} {\emph {\bibinfo {title}
  {Decoherence and the Appearance of a Classical World in Quantum Theory}}},\
  Physics and Astronomy Online Library\ (\bibinfo  {publisher} {Springer},\
  \bibinfo {year} {2003})\BibitemShut {NoStop}%
\bibitem [{\citenamefont {Schlosshauer}(2005)}]{schlosshauer2005decoherence}%
  \BibitemOpen
  \bibfield  {author} {\bibinfo {author} {\bibfnamefont {M.}~\bibnamefont
  {Schlosshauer}},\ }\href {\doibase 10.1103/RevModPhys.76.1267} {\bibfield
  {journal} {\bibinfo  {journal} {Rev. Mod. Phys.}\ }\textbf {\bibinfo {volume}
  {76}},\ \bibinfo {pages} {1267} (\bibinfo {year} {2005})}\BibitemShut
  {NoStop}%
\bibitem [{\citenamefont {Hackermüller}\ \emph {et~al.}(2003)\citenamefont
  {Hackermüller}, \citenamefont {Hornberger}, \citenamefont {Brezger},
  \citenamefont {Zeilinger},\ and\ \citenamefont
  {Arndt}}]{hackermuller2003decoherence}%
  \BibitemOpen
  \bibfield  {author} {\bibinfo {author} {\bibfnamefont {L.}~\bibnamefont
  {Hackermüller}}, \bibinfo {author} {\bibfnamefont {K.}~\bibnamefont
  {Hornberger}}, \bibinfo {author} {\bibfnamefont {B.}~\bibnamefont {Brezger}},
  \bibinfo {author} {\bibfnamefont {A.}~\bibnamefont {Zeilinger}}, \ and\
  \bibinfo {author} {\bibfnamefont {M.}~\bibnamefont {Arndt}},\ }\href
  {\doibase 10.1007/s00340-003-1312-6} {\bibfield  {journal} {\bibinfo
  {journal} {Applied Physics B}\ }\textbf {\bibinfo {volume} {77}},\ \bibinfo
  {pages} {781} (\bibinfo {year} {2003})}\BibitemShut {NoStop}%
\bibitem [{\citenamefont {{Tegmark}}(1993)}]{tegmark1993apparent}%
  \BibitemOpen
  \bibfield  {author} {\bibinfo {author} {\bibfnamefont {M.}~\bibnamefont
  {{Tegmark}}},\ }\href {\doibase 10.1007/BF00662807} {\bibfield  {journal}
  {\bibinfo  {journal} {Foundations of Physics Letters}\ }\textbf {\bibinfo
  {volume} {6}},\ \bibinfo {pages} {571} (\bibinfo {year} {1993})},\ \Eprint
  {http://arxiv.org/abs/gr-qc/9310032} {arXiv:gr-qc/9310032 [gr-qc]}
  \BibitemShut {NoStop}%
\bibitem [{\citenamefont {Nielsen}\ and\ \citenamefont
  {Chuang}(2010)}]{nielsen2002quantum}%
  \BibitemOpen
  \bibfield  {author} {\bibinfo {author} {\bibfnamefont {M.~A.}\ \bibnamefont
  {Nielsen}}\ and\ \bibinfo {author} {\bibfnamefont {I.~L.}\ \bibnamefont
  {Chuang}},\ }\href {\doibase 10.1017/CBO9780511976667} {\emph {\bibinfo
  {title} {Quantum Computation and Quantum Information: 10th Anniversary
  Edition}}}\ (\bibinfo  {publisher} {Cambridge University Press},\ \bibinfo
  {year} {2010})\BibitemShut {NoStop}%
\bibitem [{\citenamefont {Bell}(1964)}]{bell1964einstein}%
  \BibitemOpen
  \bibfield  {author} {\bibinfo {author} {\bibfnamefont {J.~S.}\ \bibnamefont
  {Bell}},\ }\href {\doibase 10.1103/PhysicsPhysiqueFizika.1.195} {\bibfield
  {journal} {\bibinfo  {journal} {Physics Physique Fizika}\ }\textbf {\bibinfo
  {volume} {1}},\ \bibinfo {pages} {195} (\bibinfo {year} {1964})}\BibitemShut
  {NoStop}%
\bibitem [{\citenamefont {Jones}(2022)}]{jones2022comment}%
  \BibitemOpen
  \bibfield  {author} {\bibinfo {author} {\bibfnamefont {B.~J.~P.}\
  \bibnamefont {Jones}},\ }\href@noop {} {\  (\bibinfo {year} {2022})},\
  \Eprint {http://arxiv.org/abs/2209.00561} {arXiv:2209.00561 [hep-ph]}
  \BibitemShut {NoStop}%
\bibitem [{\citenamefont {{Hen}}\ \emph {et~al.}(2017)\citenamefont {{Hen}},
  \citenamefont {{Miller}}, \citenamefont {{Piasetzky}},\ and\ \citenamefont
  {{Weinstein}}}]{2017RvMP...89d5002H}%
  \BibitemOpen
  \bibfield  {author} {\bibinfo {author} {\bibfnamefont {O.}~\bibnamefont
  {{Hen}}}, \bibinfo {author} {\bibfnamefont {G.~A.}\ \bibnamefont {{Miller}}},
  \bibinfo {author} {\bibfnamefont {E.}~\bibnamefont {{Piasetzky}}}, \ and\
  \bibinfo {author} {\bibfnamefont {L.~B.}\ \bibnamefont {{Weinstein}}},\
  }\href {\doibase 10.1103/RevModPhys.89.045002} {\bibfield  {journal}
  {\bibinfo  {journal} {Reviews of Modern Physics}\ }\textbf {\bibinfo {volume}
  {89}},\ \bibinfo {eid} {045002} (\bibinfo {year} {2017})},\ \Eprint
  {http://arxiv.org/abs/1611.09748} {arXiv:1611.09748 [nucl-ex]} \BibitemShut
  {NoStop}%
\bibitem [{\citenamefont {{D. Dwyer}}(2015)}]{oklo}%
  \BibitemOpen
  \bibfield  {author} {\bibinfo {author} {\bibnamefont {{D. Dwyer}}},\ }\href
  {https://github.com/dadwyer/oklo} {\enquote {\bibinfo {title} {{{OKLO}: A
  toolkit for modeling nuclides and nuclear reactions,
  https://github.com/dadwyer/oklo}},}\ } (\bibinfo {year} {2015})\BibitemShut
  {NoStop}%
\bibitem [{\citenamefont {King}\ \emph {et~al.}(2022)\citenamefont {King},
  \citenamefont {Baroni}, \citenamefont {Cirigliano}, \citenamefont {Gandolfi},
  \citenamefont {Hayen}, \citenamefont {Mereghetti}, \citenamefont {Pastore},\
  and\ \citenamefont {Piarulli}}]{king2022ab}%
  \BibitemOpen
  \bibfield  {author} {\bibinfo {author} {\bibfnamefont {G.~B.}\ \bibnamefont
  {King}}, \bibinfo {author} {\bibfnamefont {A.}~\bibnamefont {Baroni}},
  \bibinfo {author} {\bibfnamefont {V.}~\bibnamefont {Cirigliano}}, \bibinfo
  {author} {\bibfnamefont {S.}~\bibnamefont {Gandolfi}}, \bibinfo {author}
  {\bibfnamefont {L.}~\bibnamefont {Hayen}}, \bibinfo {author} {\bibfnamefont
  {E.}~\bibnamefont {Mereghetti}}, \bibinfo {author} {\bibfnamefont
  {S.}~\bibnamefont {Pastore}}, \ and\ \bibinfo {author} {\bibfnamefont
  {M.}~\bibnamefont {Piarulli}},\ }\href@noop {} {\  (\bibinfo {year}
  {2022})},\ \Eprint {http://arxiv.org/abs/2207.11179} {arXiv:2207.11179
  [nucl-th]} \BibitemShut {NoStop}%
\bibitem [{\citenamefont {Pastore}\ \emph {et~al.}(2018)\citenamefont
  {Pastore}, \citenamefont {Carlson}, \citenamefont {Cirigliano}, \citenamefont
  {Dekens}, \citenamefont {Mereghetti},\ and\ \citenamefont
  {Wiringa}}]{pastore2018neutrinoless}%
  \BibitemOpen
  \bibfield  {author} {\bibinfo {author} {\bibfnamefont {S.}~\bibnamefont
  {Pastore}}, \bibinfo {author} {\bibfnamefont {J.}~\bibnamefont {Carlson}},
  \bibinfo {author} {\bibfnamefont {V.}~\bibnamefont {Cirigliano}}, \bibinfo
  {author} {\bibfnamefont {W.}~\bibnamefont {Dekens}}, \bibinfo {author}
  {\bibfnamefont {E.}~\bibnamefont {Mereghetti}}, \ and\ \bibinfo {author}
  {\bibfnamefont {R.~B.}\ \bibnamefont {Wiringa}},\ }\href {\doibase
  10.1103/PhysRevC.97.014606} {\bibfield  {journal} {\bibinfo  {journal} {Phys.
  Rev. C}\ }\textbf {\bibinfo {volume} {97}},\ \bibinfo {pages} {014606}
  (\bibinfo {year} {2018})},\ \Eprint {http://arxiv.org/abs/1710.05026}
  {arXiv:1710.05026 [nucl-th]} \BibitemShut {NoStop}%
\bibitem [{\citenamefont {Chadwick}\ \emph {et~al.}(2011)\citenamefont
  {Chadwick}, \citenamefont {Herman}, \citenamefont {Oblozinsky}, \citenamefont
  {Dunn}, \citenamefont {Danon}, \citenamefont {Kahler}, \citenamefont {Smith},
  \citenamefont {Pritychenko}, \citenamefont {Arbanas}, \citenamefont
  {Arcilla}, \citenamefont {Brewer}, \citenamefont {Brown}, \citenamefont
  {Capote}, \citenamefont {Carlson}, \citenamefont {Cho}, \citenamefont
  {Derrien}, \citenamefont {Guber}, \citenamefont {Hale}, \citenamefont
  {Hoblit},\ and\ \citenamefont {Young}}]{CHADWICK20112887}%
  \BibitemOpen
  \bibfield  {author} {\bibinfo {author} {\bibfnamefont {M.}~\bibnamefont
  {Chadwick}}, \bibinfo {author} {\bibfnamefont {M.}~\bibnamefont {Herman}},
  \bibinfo {author} {\bibfnamefont {P.}~\bibnamefont {Oblozinsky}}, \bibinfo
  {author} {\bibfnamefont {M.}~\bibnamefont {Dunn}}, \bibinfo {author}
  {\bibfnamefont {Y.}~\bibnamefont {Danon}}, \bibinfo {author} {\bibfnamefont
  {A.}~\bibnamefont {Kahler}}, \bibinfo {author} {\bibfnamefont
  {D.}~\bibnamefont {Smith}}, \bibinfo {author} {\bibfnamefont
  {B.}~\bibnamefont {Pritychenko}}, \bibinfo {author} {\bibfnamefont
  {G.}~\bibnamefont {Arbanas}}, \bibinfo {author} {\bibfnamefont
  {R.}~\bibnamefont {Arcilla}}, \bibinfo {author} {\bibfnamefont
  {R.}~\bibnamefont {Brewer}}, \bibinfo {author} {\bibfnamefont
  {D.}~\bibnamefont {Brown}}, \bibinfo {author} {\bibfnamefont
  {R.}~\bibnamefont {Capote}}, \bibinfo {author} {\bibfnamefont
  {A.}~\bibnamefont {Carlson}}, \bibinfo {author} {\bibfnamefont {Y.-S.}\
  \bibnamefont {Cho}}, \bibinfo {author} {\bibfnamefont {H.}~\bibnamefont
  {Derrien}}, \bibinfo {author} {\bibfnamefont {K.}~\bibnamefont {Guber}},
  \bibinfo {author} {\bibfnamefont {G.}~\bibnamefont {Hale}}, \bibinfo {author}
  {\bibfnamefont {S.}~\bibnamefont {Hoblit}}, \ and\ \bibinfo {author}
  {\bibfnamefont {P.}~\bibnamefont {Young}},\ }\href {\doibase
  10.1016/j.nds.2011.11.002} {\bibfield  {journal} {\bibinfo  {journal}
  {Nuclear Data Sheets}\ }\textbf {\bibinfo {volume} {112}},\ \bibinfo {pages}
  {2887} (\bibinfo {year} {2011})}\BibitemShut {NoStop}%
\bibitem [{\citenamefont {Shibata}\ \emph {et~al.}(2011)\citenamefont
  {Shibata}, \citenamefont {Iwamoto}, \citenamefont {Nakagawa}, \citenamefont
  {Iwamoto}, \citenamefont {Ichihara}, \citenamefont {Kunieda}, \citenamefont
  {Chiba}, \citenamefont {Furutaka}, \citenamefont {Otuka}, \citenamefont
  {Ohsawa}, \citenamefont {Murata}, \citenamefont {Matsunobu}, \citenamefont
  {Zukeran}, \citenamefont {Kamada},\ and\ \citenamefont {Katakura}}]{jendl}%
  \BibitemOpen
  \bibfield  {author} {\bibinfo {author} {\bibfnamefont {K.}~\bibnamefont
  {Shibata}}, \bibinfo {author} {\bibfnamefont {O.}~\bibnamefont {Iwamoto}},
  \bibinfo {author} {\bibfnamefont {T.}~\bibnamefont {Nakagawa}}, \bibinfo
  {author} {\bibfnamefont {N.}~\bibnamefont {Iwamoto}}, \bibinfo {author}
  {\bibfnamefont {A.}~\bibnamefont {Ichihara}}, \bibinfo {author}
  {\bibfnamefont {S.}~\bibnamefont {Kunieda}}, \bibinfo {author} {\bibfnamefont
  {S.}~\bibnamefont {Chiba}}, \bibinfo {author} {\bibfnamefont
  {K.}~\bibnamefont {Furutaka}}, \bibinfo {author} {\bibfnamefont
  {N.}~\bibnamefont {Otuka}}, \bibinfo {author} {\bibfnamefont
  {T.}~\bibnamefont {Ohsawa}}, \bibinfo {author} {\bibfnamefont
  {T.}~\bibnamefont {Murata}}, \bibinfo {author} {\bibfnamefont
  {H.}~\bibnamefont {Matsunobu}}, \bibinfo {author} {\bibfnamefont
  {A.}~\bibnamefont {Zukeran}}, \bibinfo {author} {\bibfnamefont
  {S.}~\bibnamefont {Kamada}}, \ and\ \bibinfo {author} {\bibfnamefont
  {J.}~\bibnamefont {Katakura}},\ }\href {\doibase
  10.1080/18811248.2011.9711675} {\bibfield  {journal} {\bibinfo  {journal}
  {Journal of Nuclear Science and Technology}\ }\textbf {\bibinfo {volume}
  {48}},\ \bibinfo {pages} {1 } (\bibinfo {year} {2011})}\BibitemShut {NoStop}%
\bibitem [{\citenamefont {Littlejohn}\ \emph {et~al.}(2018)\citenamefont
  {Littlejohn}, \citenamefont {Conant}, \citenamefont {Dwyer}, \citenamefont
  {Erickson}, \citenamefont {Gustafson},\ and\ \citenamefont
  {Hermanek}}]{Littlejohn:2018hqm}%
  \BibitemOpen
  \bibfield  {author} {\bibinfo {author} {\bibfnamefont {B.~R.}\ \bibnamefont
  {Littlejohn}}, \bibinfo {author} {\bibfnamefont {A.}~\bibnamefont {Conant}},
  \bibinfo {author} {\bibfnamefont {D.~A.}\ \bibnamefont {Dwyer}}, \bibinfo
  {author} {\bibfnamefont {A.}~\bibnamefont {Erickson}}, \bibinfo {author}
  {\bibfnamefont {I.}~\bibnamefont {Gustafson}}, \ and\ \bibinfo {author}
  {\bibfnamefont {K.}~\bibnamefont {Hermanek}},\ }\href {\doibase
  10.1103/PhysRevD.97.073007} {\bibfield  {journal} {\bibinfo  {journal} {Phys.
  Rev. D}\ }\textbf {\bibinfo {volume} {97}},\ \bibinfo {pages} {073007}
  (\bibinfo {year} {2018})},\ \Eprint {http://arxiv.org/abs/1803.01787}
  {arXiv:1803.01787 [nucl-th]} \BibitemShut {NoStop}%
\bibitem [{\citenamefont {Vogel}\ and\ \citenamefont
  {Engel}(1989)}]{Vogel:1989iv}%
  \BibitemOpen
  \bibfield  {author} {\bibinfo {author} {\bibfnamefont {P.}~\bibnamefont
  {Vogel}}\ and\ \bibinfo {author} {\bibfnamefont {J.}~\bibnamefont {Engel}},\
  }\href {\doibase 10.1103/PhysRevD.39.3378} {\bibfield  {journal} {\bibinfo
  {journal} {Phys. Rev. D}\ }\textbf {\bibinfo {volume} {39}},\ \bibinfo
  {pages} {3378} (\bibinfo {year} {1989})}\BibitemShut {NoStop}%
\bibitem [{\citenamefont {Esteban}\ \emph {et~al.}(2020)\citenamefont
  {Esteban}, \citenamefont {Gonzalez-Garcia}, \citenamefont {Maltoni},
  \citenamefont {Schwetz},\ and\ \citenamefont {Zhou}}]{Esteban:2020cvm}%
  \BibitemOpen
  \bibfield  {author} {\bibinfo {author} {\bibfnamefont {I.}~\bibnamefont
  {Esteban}}, \bibinfo {author} {\bibfnamefont {M.~C.}\ \bibnamefont
  {Gonzalez-Garcia}}, \bibinfo {author} {\bibfnamefont {M.}~\bibnamefont
  {Maltoni}}, \bibinfo {author} {\bibfnamefont {T.}~\bibnamefont {Schwetz}}, \
  and\ \bibinfo {author} {\bibfnamefont {A.}~\bibnamefont {Zhou}},\ }\href
  {\doibase 10.1007/JHEP09(2020)178} {\bibfield  {journal} {\bibinfo  {journal}
  {JHEP}\ }\textbf {\bibinfo {volume} {09}},\ \bibinfo {pages} {178} (\bibinfo
  {year} {2020})},\ \Eprint {http://arxiv.org/abs/2007.14792} {arXiv:2007.14792
  [hep-ph]} \BibitemShut {NoStop}%
\bibitem [{\citenamefont {Akhmedov}\ and\ \citenamefont
  {Smirnov}(2022)}]{akhmedov2022damping}%
  \BibitemOpen
  \bibfield  {author} {\bibinfo {author} {\bibfnamefont {E.}~\bibnamefont
  {Akhmedov}}\ and\ \bibinfo {author} {\bibfnamefont {A.~Y.}\ \bibnamefont
  {Smirnov}},\ }\href@noop {} {\  (\bibinfo {year} {2022})},\ \Eprint
  {http://arxiv.org/abs/2208.03736} {arXiv:2208.03736 [hep-ph]} \BibitemShut
  {NoStop}%
\end{thebibliography}%

\end{document}